\documentclass[10pt,twocolumn,romanappendices]{IEEEtran-v17} 
\let\labelindent\relax % labelindent prevent from using enumitem, but is only there for legacy reasons, and hence we disable it.
\usepackage{enumitem}

\usepackage{color}

\ifCLASSINFOpdf
  \usepackage[pdftex]{graphicx}
\else  
\fi  
\usepackage[cmex10]{amsmath}
\usepackage[tight,footnotesize]{subfigure}

\usepackage[utf8]{inputenc}

\usepackage{amssymb}
\usepackage{bbm}
\usepackage{bm}
\usepackage{xspace}
\usepackage[draft,nomargin,footnote]{fixme}
\usepackage{tikz}
\usepackage{pgfplots}
\usepackage{dsfont}
\usepackage{bbold}
\pgfplotsset{compat=newest}
\pgfplotsset{plot coordinates/math parser=false}
\usetikzlibrary{patterns}
\newlength\figureheight
\newlength\figurewidth
\newlength\smallfigureheight
\newlength\smallfigurewidth
\usetikzlibrary{decorations.pathreplacing}
\usetikzlibrary{intersections}

\usetikzlibrary{shapes,arrows, positioning, fit}

\makeatletter
\def\@IEEEinterspaceratioM{0.265}
\def\@IEEEinterspaceMINratioM{0.1651}
\def\@IEEEinterspaceMAXratioM{0.38}

\def\@IEEEinterspaceratioB{0.31}
\def\@IEEEinterspaceMINratioB{0.19}
\def\@IEEEinterspaceMAXratioB{0.38}
\@IEEEtunefonts
\makeatother
\newtheorem{theorem}{Theorem}
\newtheorem{lemma}[theorem]{Lemma}
\newtheorem{corollary}[theorem]{Corollary}
\newtheorem{proposition}[theorem]{Proposition}

\providecommand{\ee}[1]{\exp\mathopen{}\left(#1\right)}

\providecommand{\tr}{^{\mathrm{T}}}
\providecommand{\indi}[1]{\mathds{1}\mathopen{}\left\{#1\right\}}

\providecommand{\tr}{^{\mathrm{T}}}

\newcommand{\E}[1]{\mathbb{E}\mathopen{}\left[#1\right]}

\newcommand{\EBigg}[1]{\mathbb{E}\mathopen{}\Bigg[#1\Bigg]}
\newcommand{\inff}[1]{\inf\mathopen{}\left\{#1\right\}}

\newcommand{\minn}[1]{\min\mathopen{}\left\{#1\right\}}
\newcommand{\maxx}[1]{\max\mathopen{}\left\{#1\right\}}
\newcommand{\EE}[2]{\mathbb{E}_{#1}\mathopen{}\left[#2\right]}

\newcommand{\farg}[1]{\mathopen{}\left( #1 \right)}

\newcommand{\pr}[1]{\text{P}\mathopen{}\left[#1\right]}

\newcommand{\vect}[1]{\boldsymbol{\mathbf{#1}}}

\newcommand{\M}{\mathrm{M}}
\newcommand{\era}{\mathrm{e}}

\def\multiset#1#2{\ensuremath{\left(\kern-.3em\left(\genfrac{}{}{0pt}{}{#1}{#2}\right)\kern-.3em\right)}}
\pgfplotsset{every x tick scale label/.style={
  at={(axis description cs:1.1,-0.1)},yshift=-0.5em,inner sep=0pt,left
 }
}

\begin{document}

%!TEX root = TCOM_bcast_v1.tex

\title{Downlink Transmission of Short Packets: Framing and Control Information Revisited}

\author{\thanks{The work of P. Popovski and K. F. Trillingsgaard was supported in part by the European Research Council (ERC Consolidator Grant Nr. 648382 WILLOW) within the Horizon 2020 Program.}Kasper Fløe Trillingsgaard\thanks{K. Trillingsgaard and P. Popovski are with the Department of Eletronic Systems, Aalborg University, 9220, Aalborg Øst, Denmark (e-mail: \{kft,petarp\}@es.aau.dk).}, \emph{Student Member, IEEE} and
Petar Popovski, \emph{Fellow, IEEE}}

\maketitle

\begin{abstract}
Cellular wireless systems rely on frame-based transmissions. The frame design is conventionally based on heuristics, consisting of a frame header and a data part. The frame header contains control information that provides pointers to the messages within the data part. In this paper, we revisit the principles of frame design and show the impact of the new design in scenarios that feature short data packets which are central to various 5G and Internet of Things applications. We treat framing for downlink transmission in an AWGN broadcast channel with $K$ users, where the sizes of the messages to the users are random variables. Using approximations from finite blocklength information theory, we establish a framework in which a message to a given user is not necessarily encoded as a single packet, but may be grouped with the messages to other users and benefit from the improved efficiency of longer codes. This requires changes in the way control information is sent, and it requires that the users need to spend power decoding other messages, thereby increasing the average power consumption. We show that the common heuristic design is only one point on a curve that represents the trade-off between latency and power consumption.
\end{abstract}

%\begin{IEEEkeywords}
%Variable-length coding, multiple-access channel
%\end{IEEEkeywords}

% \fi
\IEEEpeerreviewmaketitle

%\section{Introduction}

\section{Introduction}
Modern high-speed wireless networks heavily depend on reliable and efficient transmission of large data packets through the use of coding and information theory. The advent of machine-to-machine (M2M), vehicular-to-vehicular (V2V), and various streaming systems have spawned a renewed interest in developing information theoretical bounds and codes for communication of short packets \cite{Popovski2014}\cite{Durisi2015}\cite{FiveDisruptive}. Additionally, these applications often have tight reliability and latency constraints compared to typical wireless systems today. Communication at shorter blocklengths introduces several new challenges which are not present when considering communication of larger data packets. For example, the overhead caused by control signals and header data is insignificant if large data packets are sent, and hence, this overhead is often neglected in the analysis of protocols. However, more stringent latency requirements lead to shortened blocklengths for transmission such that 
the size of control information may approach, or even exceed, the size of the data part in the packet. This is especially true for multiuser systems such as broadcast channels, two-way channels, or multiple access channels, where the control information must include information about the packet structure, security, and user address information for identification purposes.  

The fundamentals of communication of short packets have been addressed by Strassen and, recently, Polyanskiy \emph{et al.} in \cite{Strassen2009} and \cite{Polyanskiy2010b}. It was shown that the maximum coding rate of a fixed-length code with $n$ channel uses and maximum error probability $\varepsilon$ over a discrete-time AWGN point-to-point channel has an asymptotic expansion given by
\begin{IEEEeqnarray}{rCl}
  R^*(n,\varepsilon) = C - \sqrt{\frac{V}{n}}Q^{-1}(\varepsilon) + \frac{1}{2 n}\log_2 n  +\mathcal{O}\left(\frac{1}{n}\right)\label{eq:fbl}
\end{IEEEeqnarray} 
as $n\rightarrow \infty$.
Here, $C$ is the Shannon capacity, $V$ is the channel dispersion, and $Q^{-1}(\cdot)$ denotes the inverse $Q$-function. In addition to the asymptotic expansion in \eqref{eq:fbl}, \cite{Polyanskiy2010b} used nonasymptotic bounds to numerically demonstrate that $R^*(n,\varepsilon)$ is tightly approximated by the first three terms of \eqref{eq:fbl}.
The approximation \eqref{eq:fbl} and similar ones are important in the design of communication systems because the specifics of code selection can be neglected in the optimization of protocol parameters. For example, such approximations have been applied in the optimization of packet scheduling problems \cite{Xu2016}, hybrid ARQ protocols \cite{Makki2014}, and cloud radio access networks \cite{Khalili2015}. 

In this paper, we consider downlink transmission with a discrete-time AWGN broadcast channel that consists of a transmitter and $K$ users. Downlink transmissions are organized in frames, whose structure is the main topic of this paper. In each frame, there is a message from the transmitter to the $k$-th user with a certain probability $1-q$. If, in a given frame, there is a message for user $k$, then this user is said to be \emph{active} in that frame. The size of the message to user $k$ is denoted by $D_k$ and is a random variable itself. Hence, the transmitter needs to convey information about which users are active, the structure of the transmission, and sizes of the messages. As a result, the \emph{frame duration}, which corresponds to the total transmission time, and the total power consumption at the users are also random variables. An important observation from \eqref{eq:fbl} is that larger data packets are encoded more efficiently. This introduces an interesting trade-off with two extremes: (a) in a broadcast setting one can either encode all messages in one large packet, or (b) one can encode each message separately, which is the norm in modern wireless protocols. In (a), the average frame duration is minimized, which implies that the average \emph{latency} across the users is minimized. However, the downside of (a) is that all users need to receive for the whole period of transmission to be able to decode their messages, which is undesirable for devices that are power-constrained. The latter approach (b), depicted in Fig.~\ref{fig:protocol_conventional}, uses codes which are less efficient, and thus the average frame duration is larger. On the other hand, each user only needs to decode the information intended for that user. The key point, however, is that these design considerations enlarge the space of feasible protocols and enable the protocol designer to seek a trade-off between frame duration (latency) and power consumption at the users. Despite this trade-off, practically all wireless systems solely use the extreme approach (b). 

\subsubsection*{Contribution}
The purpose of this paper is to revisit the way a downlink frame is designed when it contains short packets. Specifically, it aims at exploring the trade-off between the average frame duration and the average power consumption at the users. 
Instead of using a traditional frame structure, we enlarge the design space for a frame by doing the following: the users are divided into groups that may depend on the realization of the message sizes and the messages of each group are jointly encoded using optimal channel codes. We analyze the problem using asymptotic expansions similar to \eqref{eq:fbl}, and we find a lower bound for the trade-off curve. Next, we introduce three protocols: (a) a genie-aided protocol with performance close to the lower bound, (b) protocol with a fixed message that works for the case in which each message has either the size $0$ or $\alpha\in\mathbb{N}$ bits, and (c) a protocol with variable message sizes, where the message sizes are distributed according to a probability mass function $P_D$ with finite and nonnegative integer support. The protocols (b) and (c) both convey enough control information to make them practically usable. Our numerical results demonstrate trade-offs which are particularly interesting when the message sizes are small. 
%An essential message of this paper is that one can achieve better overall system performance by allowing the users to decode the messages destined to other users. 

\subsubsection*{Organization} Section~\ref{sec:fbl} introduces the finite blocklength approximations and bounds for optimal channel codes while the system model is introduced in Section~\ref{sec:system_model}. Section~\ref{sec:lower_bnd} presents a lower bound for the average power at each user expressed as a function of the average frame duration. Section~\ref{sec:protocol_design} provides some concrete protocol designs, which are subsequently compared with the lower bound. Finally, numerical examples are presented in Section~\ref{sec:numerical} and Section~\ref{sec:conclusions} concludes the paper.

\subsubsection*{Notation}
 Vectors are denoted by boldface letters (e.g., $\vect{x}$) while their entries are denoted by roman letters (e.g., $x_i$). We denote the $n$-dimensional all-zero vector and all-one vector by $\vect{0}_n$ and $\vect{1}_n$, respectively. We denote by $\vect{\bar 0}^n_i(x)$ the $n$-dimensional vector with $x$ in the $i$-th entry and zeroes in the rest.
We let $\oplus$ denote the concatenation of two bit string, e.g., for $\vect{a} \in\{0,1\}^n$ and $\vect{b}\in\{0,1\}^m$, $\vect{a}\oplus \vect{b}$ is the concatenated bit string. 
Throughout the paper, the index $k$ belongs always to the set $\mathcal{K}\triangleq \{1,\cdots,K\}$, although this is sometimes not explicitly mentioned.  We define the upper concave envelope of a function $f: \mathbb{R}_+ \mapsto \mathbb{R}_+$ as $\text{uce}(f) \triangleq \inf_g\{g \geq f \text{ and } g \text{ is concave}\}$. Similarly, the lower convex envelope is defined by $\text{lce}(f) \triangleq \sup_g\{g \leq f \text{ and } g \text{ is convex}\}$. Finally, $\mathbb{N}$ denotes the set of positive integers, $\mathbb{Z}_+ \triangleq \mathbb{N}\cup \{0\}$, and the symbol $\mathbb{R}$ indicate the set of real numbers.

\begin{figure}[!t]
  \centering
  \includegraphics[width=3.2in]{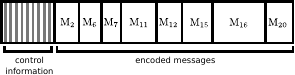}
  \caption{Conventional approach to downlink broadcasting. We denote by $\M_2,\M_6,\cdots,\M_{20}$ the messages of varying size (in bits) destined to the active users. An initial packet contains control information that defines the structure of the remaining part of the transmission. Each message is encoded separately.}
  \label{fig:protocol_conventional}
\end{figure}

%This paper is organized as follows. The next section describes the system model. Section~\ref{sec:fbl} briefly addresses the approximations of the finite blocklength bounds, Section~\ref{sec:protocol_design} discusses design considerations for protocol design for downlink broadcast for short packets and describes our proposed protocol. Finally, we evaluate the proposed protocol in Section~\ref{sec:numerical} and conclude the paper in Section~\ref{sec:conclusions}.

\section{Finite Blocklength Bounds and Approximations}
\label{sec:fbl}
In our analysis, we apply results from finite blocklength information theory. 
For the (real) AWGN channel under a short-term power constraint $P$, \cite{Strassen2009} and \cite{Polyanskiy2010b} showed that the maximum coding rate $R^*(n,\varepsilon)$ of a code with fixed blocklength $n$ and error probability   $\varepsilon\in(0,1)$ has the asymptotic expansion given by \eqref{eq:fbl}, where the channel capacity $C$ and the channel dispersion $V$ are given by
\begin{IEEEeqnarray}{rCl}
  C &\triangleq& \frac{1}{2} \log_2(1+P)\label{eq:capacity}
  \end{IEEEeqnarray}
  and
  \begin{IEEEeqnarray}{rCl}
V &\triangleq& \frac{P(P+2)}{2(P+1)^2} \log_2(\exp(1))^2\label{eq:dispersion}
\end{IEEEeqnarray}
respectively. One can obtain tight nonasymptotic upper and lower bounds for $R^*(n,\varepsilon)$ using the achievability and converse bounds in \cite{Polyanskiy2010b}, and  it was numerically demonstrated  that the first three terms of the right-hand side of  \eqref{eq:fbl} provide a tight approximation of $R^*(n,\varepsilon)$.
%Recently, Yang \emph{et al.} showed in \cite{Wei2015} that the maximum coding rate under a long-term power constraint, i.e., the average power over all codewords should satisfy the power constaint $P$, is also well-approximated by the first three terms of \eqref{eq:fbl}, provided that the blocklength and the error probability are small. As an example, it was shown that the normal approximation \eqref{eq:fbl} provides a good approximation for $P=0\ \text{dB}$, $\varepsilon = 10^{-3}$, and $n \leq 3 \times 10^5$. For longer blocklength and lower error probabilities, the authors find a different approximation \cite[Eq.~(8)]{Wei2015}.

\begin{figure}[!t] 
\centering
\begin{tikzpicture}[thick,scale=0.95, every node/.style={scale=0.95}]
\begin{axis}[
grid=both,
tick align=center,
width=9.1cm, 
height=7.1cm,
minor tick num=4,
every minor grid/.style={opacity=0.2},
xmin=0,
xtick={0,500,1000,1500,2000},
xticklabels={{   0},{ 500},{1000}, {1500},{2000}},
xmax=2000,
ymin=1.2,
ymax=1.8,
legend entries ={ {}{$N_a(k,\varepsilon)/k$},{}{$\bar N(k,\varepsilon)/k$},{}{$N(k,\varepsilon)/k$},{}{$N_c(k,\varepsilon)/k$},{}{$1/C$}},  
xlabel={Information bits $k$},
ylabel={Ch. uses/information bit}
]
\addplot [
    name path=ach,
    color=magenta,
    solid,  
    line width=1.0pt 
    ] table [x index = {0}, y index = {1}, col sep=comma]{Nach.dat}; 
\addplot [
    name path=approx2,
    color=green,  
    solid,  
    line width=1.0pt 
    ] table [x index = {0}, y index = {3}, col sep=comma]{Napproximations.dat};
\addplot [
    name path=approx,
    color=red,
    solid,  
    line width=1.0pt 
    ] table [x index = {0}, y index = {2}, col sep=comma]{Napproximations.dat};
\addplot [
    name path=converse,
    color=blue,
    dashed,  
    line width=1.0pt 
    ] table [x index = {0}, y index = {1}, col sep=comma]{Nconverse.dat};

\addplot [
    name path=cap,
    color=black,
    solid,  
    line width=1.0pt 
    ] table [x index = {0}, y index = {1}, col sep=comma]{Napproximations.dat};
\end{axis}
\end{tikzpicture} 
\caption{Bounds and approximations for $N^*(k,\varepsilon)$ plotted for $\varepsilon=10^{-3}$ and $P=0\ \text{dB}$. The converse $N_c(k,\varepsilon)$ and achievability bound $N_a(k,\varepsilon)$ are plotted using the SPECTRE toolbox.}  
\label{fig:fbl}
\end{figure}
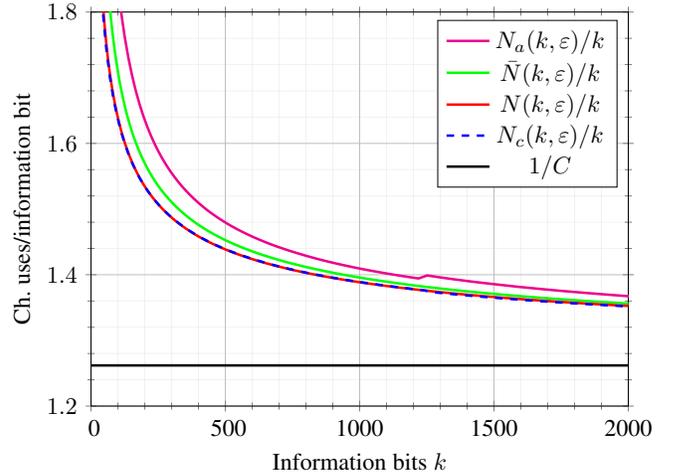

We define $N^*(k, \varepsilon) \triangleq \minn{n \geq 0: n R^*(n,\varepsilon) \geq k}$ for $k\geq 1$ and $N^*(0,\varepsilon) \triangleq  0$ which is the smallest number of channel uses that allows the encoding of $k$ bits with error probability $\varepsilon$.
We obtain the following asymptotic approximation of $N^*(k, \varepsilon)$ as $k\rightarrow \infty$:
\begin{IEEEeqnarray}{rCl}
  N^*(k, \varepsilon) &=& \frac{k}{C} + \sqrt{\frac{V k}{C^3}}Q^{-1}(\varepsilon) - \frac{1}{2 C} \log_2\frac{k}{C} + \mathcal{O}(1).\IEEEeqnarraynumspace\label{eq:Nasymptotic}
\end{IEEEeqnarray}
This can be verified by setting $\bar n$ equal to RHS of \eqref{eq:Nasymptotic} and by computing $\bar n R^*(\bar n,\varepsilon)$. Then, one finds that $\bar n R^*(\bar n,\varepsilon) = k +\mathcal{O}(1)$ from which \eqref{eq:Nasymptotic} follows.
We define the approximation 
\begin{IEEEeqnarray}{rCl}
  N(k,\varepsilon) \triangleq \text{uce}\mathopen{}\left(\frac{k}{C} + \sqrt{\frac{V k}{C^3}}Q^{-1}(\varepsilon) - \frac{1}{2 C} \log_2\frac{k}{C} \right)\label{eq:Napprox_uce}
\end{IEEEeqnarray}
where $\text{uce}(\cdot)$ stands for the upper concave envelope. It can be shown that the approximation of $N^*(k,\varepsilon)$ inside $\text{uce}(\cdot)$ in \eqref{eq:Napprox_uce} is concave for $k \geq \frac{4 C}{Q^{-1}(\varepsilon)^2 V \log_e(2)^2}$, implying that $N(k,\varepsilon) = N^*(k,\varepsilon) + \mathcal{O}(1)$. Additionally, in all numerical examples in this paper, we have $\frac{4 C}{Q^{-1}(\varepsilon)^2 V \log_e(2)^2} < 1$, and hence the upper concave envelope does not affect our numerical results. In Fig.~\ref{fig:fbl}, we have plotted the $\kappa\beta$-achievability bound $N_a(k,\varepsilon)$ and the meta-converse bound $N_c(k,\varepsilon)$ from \cite{Polyanskiy2010b} along with the approximation \eqref{eq:Napprox_uce}, and $\bar N(k,\varepsilon)=n$, where $n$ is a solution to:
\begin{IEEEeqnarray}{rCl}
  n C - \sqrt{n V}Q^{-1}(\varepsilon) + \frac{1}{2} \log_2 n = k.
\end{IEEEeqnarray}
We observe that $N(k,\varepsilon)$ provides an approximation of $N^*(k,\varepsilon)$ that matches the converse bound closely. 
In the remaining part of this paper, when referring to the blocklength of an optimal code conveying $k$ bits with a probability of error not exceeding $\varepsilon$, we consistently use the approximation $N(k,\varepsilon)$ in place of $N^*(k,\varepsilon)$ in all computations and derivations.

\begin{comment}
\begin{IEEEproof}
We set 
\begin{IEEEeqnarray}{rCl}
\bar n = \frac{k}{C} + \sqrt{\frac{V k}{C^3}}Q^{-1}(\epsilon) - \frac{1}{2 C} \log_2\frac{k}{C} + \frac{d}{C}
\end{IEEEeqnarray}
  and compute $\bar n R^*(\bar n,\epsilon)$:

  \begin{IEEEeqnarray}{rCl}
    \bar n R^*(\bar n,\epsilon) &=& k + \sqrt{\frac{V k}{C}}Q^{-1}(\epsilon)- \frac{1}{2} \log_2\frac{k}{C}  + d \nonumber\\
    && {}-\sqrt{\frac{V k}{C}}Q^{-1}(\epsilon)  +\frac{1}{2} \log_2\farg{\frac{k}{C}}+ \mathcal{O}(1).
  \end{IEEEeqnarray}
  Here, we have used that $\sqrt{Vk/C + \mathcal{O}(\sqrt{k})} = \sqrt{V k/C}+\mathcal{O}(1)$ and that $\log_2(k/C + \mathcal{O}(\sqrt{k})) = \log_2(k/C)+\mathcal{O}(1/\sqrt{k})$. This completes the proof
\end{IEEEproof}
\end{comment}

\section{System model}\label{sec:system_model}
We consider an AWGN broadcast channel with one transmitter and $K$ users. In the $t$-th time slot, the $k$-th user receive
\begin{IEEEeqnarray}{rCl}
Y_{k,t} \triangleq \sqrt{\gamma_k} X_t + Z_{k,t}.\label{eq:system_model}
\end{IEEEeqnarray}
where $Z_{k,t}\sim \mathcal{N}(0,1)$ and $X_t\in\mathbb{R}$ is the channel input. Throughout the paper, we assume that $\gamma_k = 1$. The assumption of equal channel conditions can, to some extend, be justified as follows. Consider a downlink broadcast scenario with many users with varying channel conditions. A viable communication strategy is to first divide the users into several CSI-groups such that the users assigned to a certain CSI-group have similar channel conditions. Then, the transmitter serves each CSI-group sequentially, and our system model in \eqref{eq:system_model} models a single CSI-group. A satellite-based broadcast system with line-of-sight to all users and predictable channel conditions constitute a practical example of our system model. If, however, CSI-grouping is not performed, then the transmitter needs to protect a packet destined to multiple users with a code that is strong enough to ensure that even the worst-channel user can decode. The assumption of nonfading channels is mainly introduced for simplicity, but we note that there are results in finite blocklength information theory for fading channels \cite{Yang2014}. 

The message $\M_k$ destined to the $k$-th user is nonempty with probability $1-q\in(0,1)$, and we say that the $k$-th user is \emph{active} if there is a message destined to that user. We assume that the size of the message $\M_k$ (in bits) is given by $D_k\in\mathbb{Z}_+$ which is a discrete random variable distributed independently according to the probability mass function
\begin{IEEEeqnarray}{rCl}
  P_D(d) \triangleq \left\{ \begin{array}{ll}
   q & \text{if } d=0  \\
   (1-q)p_i & \text{if } d=\alpha_i \text{  for  } i \in\{1,\cdots,S\}.
  \end{array}
    \right.
\end{IEEEeqnarray}
The message $\mathrm{M}_k$ is drawn uniformly randomly from the set $\{0,1\}^{D_k}$.  We use $\vect{\alpha}=(\alpha_1,\alpha_2, \ldots, \alpha_S)$ to denote a $S$-dimensional vector of distinct ordered positive integers ($\alpha_i<\alpha_s$ if $i<s$) that correspond to the possible message sizes. 

%For simplicity, we assume that the sizes are of the form $D_k \triangleq a J$ for a random variable $J \in \{0,\cdots, S\}$ and a positive integer $a\in\mathbb{N}$.

The frame duration $T$ is a random variable that depends on the message sizes $\{D_k\}$. The transmitter encodes the message $\{\M_k\}$ into a sequence of channel inputs using the encoder function $f_t(\cdot)$ such that
\begin{align}
  X_t \triangleq f_t(\{D_k\},\{\M_k\})
\end{align}
for $t\in\{1,\cdots,T\}$ and $X_t = 0$ for $t\in\{T+1,\cdots\}$. Additionally, we require that 
\begin{IEEEeqnarray}{rCl}
  \E{\frac{1}{T}\sum_{i=1}^T X_t} \leq P.
\end{IEEEeqnarray}

We define the ON-OFF function $g_{k,t}: (\mathbb{R} \cup \{\era\})^{t-1}\rightarrow \{0,1\}$ that defines the receiver activity for user $k$: 
\begin{align}
  \bar Y_{k,t} \triangleq \left\{ \begin{array}{ll}
  Y_{k,t}, &  g_{k,t}(\bar Y_{k}^{t-1}) = 1, \textrm{receiver is ON} \\
  \era, & \text{receiver is OFF}
  \end{array}
  \right..
\end{align}
\begin{sloppypar}
The ON-OFF function replaces the $t$-th channel output with an erasure if the user is OFF at that time. The \emph{stopping time} $T_k$ represents the time index of the last nonerased channel output in the sequence $\bar Y_{k,t}$; after $T_k$ the receiver $k$ is OFF until the end of the frame. Formally, $T_k \triangleq \inff{n \geq 1:\forall t > n,  g_{k,t}(\bar Y_{k}^{t-1}) = 0}$ for which we require $T_k < \infty$. Considering that a user can only use the channel outputs for which it is ON, we define the decoding function $h_{k,t}(\bar Y_k^t)$ to estimate the message $\M_k$ based on $\bar Y_{k}^t$. The ON-OFF functions are causal in the sense that the decision of whether the users are ON at time $t$ depends on previous channel outputs, $\bar Y_{k}^{t-1}$. Unless an error occurs during decoding, the stopping times $T_k$ are less than or equal $T$ for any practical applications of this model. We merely define $T_k$ to emphasize that $T$ is a random variable which is not known by the users, and hence the users need to obtain this information through the sequence $\bar Y_{k,t}$. In a conventional approach to downlink broadcast, as depicted in Fig.~\ref{fig:protocol_conventional},  control information in the initial packet defines the structure of the remaining transmission. Hence, after successfully decoding the control information in the initial packet, the $k$-th user knows $T_k$ and when to be ON and OFF to receive the message intended for that user.
\end{sloppypar}

The average power consumption of the $k$-th user is given by
\begin{IEEEeqnarray}{rCl}
P_k \triangleq \E{\sum_{i=1}^{T_k} \indi{g_{k,i}(\bar Y_{k}^{i-1} )=1}}\label{eq:power_def}
\end{IEEEeqnarray}
where $\indi{\cdot}$ is the indicator function, and is determined by the ON-OFF function. Note that $\E{P_1} = \E{P_k}$, for $k\in\{1,\cdots,K\}$, since the message sizes $D_k$ are distributed identically.
Finally, the active users need to decode their messages with reliability larger than or equal $1-\epsilon$ such that
\begin{align}
  \pr{h_{k,T_k}(\bar Y_k^{T_k}) \not= \M_k | D_k > 0} \leq \epsilon\label{eq:reliability}
\end{align}
for $k\in\{1,\cdots,K\}$ and $\epsilon\in(0,1)$. 
 
The above system model provides a general framework for the problem of downlink broadcast framing. For tractability, we constrain ourselves to an important and practical class of protocols described as follows.  The transmitter forms $L\in\mathbb{Z}_+$ packets which are encoded using optimal codes with error probabilities $\{\bar \epsilon_l\}_{l\in\{1,\cdots,L\}}$. Here, $L$ and $\{\bar \epsilon_l\}$ are random variables that depend only on $\{D_k\}$. Let $L_{\text{max}}$ be a constant that denotes the maximum number of packets that the transmitter can send, defined as the smallest integer such that $L_{\text{max}}\geq L$ for all realizations of $\{D_k\}$. Let $\{\M^{(C)}_l\}_{l=1}^{L_{\text{max}}}$ denote the control information that needs to be conveyed in order to describe how the data for different users is conveyed (see the example below). Let   
$\{D^{(C)}_l\}_{l=1}^{L_{\text{max}}}$ denote the sizes (in bits) of  $\{\M^{(C)}_l\}_{l=1}^{L_{\text{max}}}$, i.e., $\M^{(C)}_l\in\{0,1\}^{D^{(C)}_l}$ and $D_l^{(C)} = 0$ for $l> L$. Let $\{\mathcal{U}_l\}_{l=1}^{L_{\text{max}}}$ denote disjoint random sets that depend only on $\{D_k\}$ such that $\bigcup_{i=1}^{L_{\text{max}}} \mathcal{U}_l = \mathcal{K}$ and such that $\mathcal{U}_l = \emptyset$ for $l> L$. The $l$-th packet then consists of the information bits $\M^{(C)}_l \oplus \bigoplus_{k\in\mathcal{U}_l} \M_k$ which are encoded by an optimal code with reliability $\bar \epsilon_l$ using $N\farg{D^{(C)}_l+\sum_{k\in\mathcal{U}_l} D_k,\bar \epsilon_l}$ channel uses. The encoder function $f_t(\cdot,\cdot)$ is defined by sequentially transmitting the $L$ encoded packets. The frame duration $T$ is given by $\sum_{l=1}^L N\farg{D^{(C)}_l+\sum_{k\in\mathcal{U}_l} D_k,\bar \epsilon_l}$.

We assume that the optimal code has the following property: If $j$ bits are encoded into $n$ channel uses by an optimal code with error probability $\varepsilon$, then the user needs to receive all $n$ channel uses so as to decode any of the $j$ bits with error probability  $\varepsilon$.

As an illustration, we describe how the general framework is instantiated to describe a 
conventional downlink frame from Fig.~\ref{fig:protocol_conventional}. Suppose $S = 3$ such that $D_k \in \{0,\alpha_1,\alpha_2,\alpha_3\}$. As there are four possible lengths, the control information about $\{D_k\}$ can be represented by at most $2K$ information bits which are conveyed in the first packet, commonly referred to as the \emph{header}. 
We let $D^{(C)}_1 = 2K$ and let $\M_1^{(C)}$ be the bitstring of length $2K$ representing $\{D_k\}$. 
Since there is a header packet and at most $K$ other packets, we set $L_{\text{max}} = L = K+1$.
We also set $\bar \epsilon_1 = \varepsilon_1$ and $\bar \epsilon_l = \varepsilon_2$ for $l\in\{2,\cdots,L_{\text{max}}\}$ where $(\varepsilon_1,\varepsilon_2)\in[0,1]^2$ are such that $\epsilon  = 1-(1-\varepsilon_1)(1-\varepsilon_2)$. Since all control information is concentrated in the frame header, we have $D_l^{(C)}=0$ for $l\geq 2$.
The sets $\{\mathcal{U}_l\}_{l=1}^{L_{\text{max}}}$ are defined such that the header has no user data and 
$\mathcal{U}_1 = \emptyset$, while $\mathcal{U}_l = \{l-1\}$ for $l\in\{2,\cdots,L_{\text{max}}\}$. 
User $k$ is ON during the transmission of the first packet which it decodes with probability $1-\varepsilon_1$. If user $k$ successfully decodes the first packet, it learns $\{D_k\}$, and thereby it obtains a pointer to the location of the $(k+1)$-th packet, which contains the desired message $\M_k$. After decoding the header, the $k$-th user is OFF for the remaining time except when the $(k+1)$-th packet is transmitted. The $(k+1)$-th packet is successfully decoded with probability $1-\varepsilon_2$. The overall probability of error for the protocol from the viewpoint of a single user is given by $1-(1-\varepsilon_1)(1-\varepsilon_2) = \epsilon$ as desired.

For large message sizes $\alpha_s \gg 1$ we get the lower bounds:
\begin{IEEEeqnarray}{rCl}
  \E{T} \geq  \frac{K \E{D_1} }{C}\label{eq:Tbound} \\
  \E{P_1} \geq \frac{\E{D_1}}{C} .\label{eq:Pbound} 
\end{IEEEeqnarray}
When $\alpha_s \gg 1$, the control information becomes negligible, and hence for the conventional approach both $\E{T}$ and $\E{P_1}$ simultaneously approach the lower bounds in \eqref{eq:Tbound} and \eqref{eq:Pbound}. 

Our objective is to explore trade-offs between the competing goals of minimizing  $\E{T}$ and $\E{P_1}$. 

\section{Lower bound}\label{sec:lower_bnd}
We establish a lower bound by assuming that the users are provided with control information from a genie, i.e.,  $\{D_k\}$ are known at all users. In that case, the transmitter and all users can agree on a protocol that only conveys the messages $\{M_k\}$, i.e., $D^{(C)}_l = 0$ for $l\in\{1,\cdots,L_{\text{max}}\}$.  Hence, the transmitter may encode the messages $\{\mathrm{M}_k\}$ into at most $K$ separate packets such that each message is encoded in exactly one of these packets. Each packet may contain either no messages at all, a single message, or multiple concatenated messages, and they are encoded using optimal codes with error probabilities that do not exceed $\epsilon$ upon decoding; recall that all users experience the same error probability since $\gamma=1$. 
Any genie-aided protocol can be characterized using $K$ random nonnegative integer vectors $\vect{N}_l\in \mathbb{Z}_+^S$, for $l\in\{1,\cdots,K\}$, that depend only on $\{D_k\}$. The content of the $l$-th packet is described by $\vect{N}_l$; the packet encodes $N_{l,1}$ messages of length $\alpha_1$, it encodes $N_{l,2}$ messages of length $\alpha_2$, etc. Note that the integer vectors $\{\vect{N}_l\}$ do not uniquely describe which messages are encoded in which packets. For a genie-aided protocol defined by a set of vectors $\{\vect{N}_l\}$, we compute the frame duration and average power as follows
\begin{IEEEeqnarray}{rCl}
  T &=& \sum_{l=1}^K N(\vect{\alpha}\tr \vect{N}_l, \epsilon)\label{eq:prot_T} \\
  \frac{1}{K}\sum_{k=1}^K P_i &=&  \frac{1}{K}\sum_{l=1}^K \vect{1}_S\tr \vect{N}_l N(\vect{\alpha}\tr \vect{N}_l, \epsilon).\label{eq:prot_P}
\end{IEEEeqnarray}
Here, $T$ and $\frac{1}{K}\sum_{k=1}^K P_i$ are random variables that depend only on the realization of $\{D_k\}$. We aim to lower bound $\E{T} + \beta \E{P_1}$ for any $\beta>0$ and thereby obtain a lower bound on the average power consumption $\E{P_1}$ as a function of average frame duration $\E{T}$.

\begin{figure}[!t]
\centering
\begin{tikzpicture}
\begin{axis}[
/pgf/number format/.cd, 1000 sep={},
grid=both,
tick align=center,
width=0.4\paperwidth,
height=0.35\paperwidth,
ymin=2000,
xmin=8500, 
ymax=9000,
xmax=9200, 
xlabel={Frame duration [ch. uses]},
ylabel={Power [ch. uses]},
legend entries = { {}{\eqref{eq:ex4} for all $\beta>0$}, {}{\eqref{eq:ex4} with $\beta = 0.035$}, {}{\eqref{eq:ex4} with $\beta = 0.095$}, {}{\eqref{eq:ex4} with $\beta = 0.245$}}
] 
\addplot [
    color=black,
    line width=1.0pt] table [x index = {0}, y index = {1}, col sep=comma]{low_bnd_ex.dat};
    \addplot [
    color=green,
    line width=1.0pt] table [x index = {0}, y index = {2}, col sep=comma]{low_bnd_ex.dat};
    \addplot [
    color=red,
    line width=1.0pt] table [x index = {0}, y index = {3}, col sep=comma]{low_bnd_ex.dat};
    \addplot [
    color=blue, 
    line width=1.0pt] table [x index = {0}, y index = {4}, col sep=comma]{low_bnd_ex.dat};
    \addplot [
    only marks,
    mark = *,
    line width=1.0pt] table [x index = {0}, y index = {1}, col sep=comma]{low_bnd_ex_points.dat};
\end{axis}
\end{tikzpicture}
\caption{Depicts the lower bound in \eqref{eq:ex4} for three different values of $\beta$ for $P=0\ \text{dB}$, $\epsilon = 10^{-4}$, and $\alpha_1 = 1000$. The black curve is obtained by evaluating \eqref{eq:ex4} for all $\beta>0$ and by combining the resulting lower bounds. The dots correspond to five genie-aided protocols described by $(N_1,\cdots,N_4) \in [(4,0,0,0), (3, 1,0,0),(2,2,0,0),(2,1,1,0),(1,1,1,1)]$ (enumerated from top-left corner to bottom-right corner).}
\label{fig:low_bnd_ex}
\end{figure}
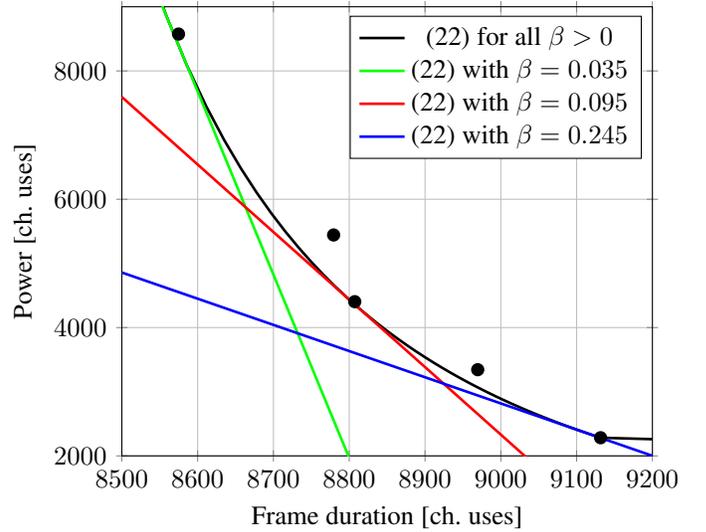
Before stating the lower bound, we introduce the technique through an example. Suppose $K=4$, $q=0$, $S=1$ such that $D_1 = \cdots= D_4 = \alpha_1$ and the frame duration and average power are deterministic. Since the users know $\{D_k\}_{k=1}^4$ and each of the four messages belongs to one encoded packet, any genie-aided protocol can be described through the four nonnegative integers  $N_1,N_2, N_3$, and $N_4$ satisfying $N_1 +\cdots+ N_4 = 4$. These integers represent the number of messages encoded in the first, second, third, and fourth packet, respectively. For fixed $\beta>0$, our objective is to minimize $T + \frac{\beta}{4} \sum_{l=1}^4 P_k$ with respect to $N_1, \cdots,N_4\in \{0,\cdots,4\}$ subject to $N_1+\cdots+N_4 = 4$. For this particular example, one can easily solve the resulting integer optimization problem. However, we can also find a lower bound on $T + \frac{1}{4}\beta \sum_{l=1}^4 P_k$ through the following steps
\begin{IEEEeqnarray}{rCl}
 \IEEEeqnarraymulticol{3}{l}{T + \frac{\beta}{4} \sum_{l=1}^4 P_k}\nonumber\\
 \quad &=& \sum_{l=1}^4 N(\alpha_1 N_l, \epsilon)+ \frac{\beta}{4}\sum_{l=1}^4  N_l N(\alpha_1 N_l, \epsilon)\\
  &\geq& \min_{\substack{n_1,\cdots,n_4\in \{0,\cdots,4\}:\\ n_1+\cdots+n_4=4  }}\sum_{l=1}^4 \Phi_\beta(n_l)\label{eq:ex1}\\
    &\geq& \min_{\substack{n_1,\cdots,n_4\in \{0,\cdots,4\}:\\ n_1+\cdots+n_4=4  }}\sum_{l=1}^4 \breve{\Phi}_\beta(n_l)\label{eq:ex2}\\
    &\geq& \min_{\substack{n_1,\cdots,n_4\in \{0,\cdots,4\}:\\ n_1+\cdots+n_4=4  }} 4\breve{\Phi}_\beta\farg{\sum_{l=1}^4 \frac{n_l}{4}}\label{eq:ex3}\\
        &=& 4\breve{\Phi}_\beta\farg{1}.\label{eq:ex4}
\end{IEEEeqnarray}
Here, \eqref{eq:ex1} follows by defining
\begin{IEEEeqnarray}{rCl}
\Phi_\beta(x) \triangleq N(\alpha_1 x, \epsilon)\left(1+ \frac{\beta x}{4}\right),
\end{IEEEeqnarray}
and by a minimization with respect to $n_1,\cdots,n_4$, \eqref{eq:ex2} follows by defining $\breve{\Phi}_\beta(\cdot)$ as the lower convex envelope of $\Phi_\beta(\cdot)$, and \eqref{eq:ex3} is by convexity of the lower convex envelope of $\Phi_\beta(\cdot)$. Interestingly, the bound in \eqref{eq:ex4} is fairly tight and simple to compute. We illustrate the bound \eqref{eq:ex4} in Fig.~\ref{fig:low_bnd_ex}, confirming the intuition that when $S=1$, one should attempt to have an equal number of non-empty messages in each packet, which for this example is $1$, $2$, or $4$ messages in each packet. 

For the general setting with arbitrary $S\geq 1$ and $K$, we apply the above ideas in the following proposition which enables us to compute a lower bound on $\E{P_1}$ for certain $\E{T}$. 
\begin{proposition}
\label{prop:prop_converse}
  For every $\beta> 0$, we have
  \begin{IEEEeqnarray}{rCl}
   \E{T} + \beta \E{P_1} \geq \E{ \vect{1}_S\tr \vect{L}_{1:S} \breve{\phi}_{\beta}\farg{ \frac{\vect{L}_{1:S}}{ \vect{1}_S\tr \vect{L}_{1:S}} }  }
  \end{IEEEeqnarray}
  where $\vect{L}\in\mathbb{Z}_+^{S+1}$ is multinomial distributed with $S+1$ categories, $K$ trials, and event probabilities $[(1-q)p_1, \cdots, (1-q)p_S,q]$, $\vect{L}_{1:S}$ denotes the first $S$ entries of $\vect{L}$, and $\breve{\phi}_{\beta}: \mathbb{R}_+^{S} \mapsto \mathbb{R}_+$ is the lower convex envelope of the function
  \begin{IEEEeqnarray}{rCl}
    \phi_\beta(\vect{x})\triangleq N\farg{\vect{\alpha}\tr \vect{x},\epsilon} \left(1+\frac{\beta \vect{1}_S\tr \vect{x}}{K}\right)\IEEEeqnarraynumspace\label{eq:non_phi_def}
  \end{IEEEeqnarray}   
  defined for $\vect{x}\in\mathbb{R}_+^{S}$.
\end{proposition}
\begin{IEEEproof}
Fix $\beta> 0$. The users $\mathcal{K}$ can be decomposed into $S+1$ disjoint subsets $\{\mathcal{U}^{(s)}\}_{s\in\{0,\cdots,S\}}$ such that $\mathcal{U}^{(s)}\triangleq \{k \in \mathcal{K}: D_k=\alpha_s\}$ for $s\in\{0,\cdots,S\}$, where we let $\alpha_0 \triangleq 0$ for notational convenience. We denote the (random) set of active users by $\mathcal{U}=\bigcup_{s=1}^S \mathcal{U}^{(s)}$.
Fix a genie-aided protocol. Then, since we assume that the users are provided with control information by a genie, the protocol must decompose the set of active users $\mathcal{U}$ into at most $K$ (possibly empty) disjoint subsets $\{\mathcal{U}_l\}_{l\in\{1,\cdots,K\}}$. Note that these subsets are random, depend only on $\{D_k\}$, and are induced by the protocol. Define the random integer vectors $\vect{N}_l\in\mathbb{Z}_+^S$, for $l\in\{1,\cdots,K\}$ and $s\in\{1,\cdots,S\}$, as follows:
\begin{IEEEeqnarray}{rCl}
N_{l,s} \triangleq \sum_{k\in\mathcal{K}} \indi{k \in\mathcal{U}_l \text{ and } D_k = \alpha_s}.
\end{IEEEeqnarray}
The average frame duration and the average power for the genie-aided protocol in terms of $\{\vect{N}_i\}$ are now given by
\begin{IEEEeqnarray}{rCl}
  \E{T} &=&  \E{\sum_{l=1}^K N(\vect{\alpha}\tr \vect{N}_l,\epsilon)}\label{eq:low_bnd_ET}\\
  \E{P_1} &=&  \E{\frac{1}{K} \sum_{l=1}^K \vect{1}_S \tr \vect{N}_l N(\vect{\alpha}\tr \vect{N}_l,\epsilon)}.\label{eq:low_bnd_EP}
\end{IEEEeqnarray}
Now, we compute a lower bound on $\E{T}+\beta \E{P_1}$ based on \eqref{eq:low_bnd_ET} and \eqref{eq:low_bnd_EP}:
\begin{IEEEeqnarray}{rCl}
  \IEEEeqnarraymulticol{3}{l}{\E{T}+\beta \E{P_1}}  \nonumber\\\quad
  &=& \E{\sum_{l=1}^K N(\vect{\alpha}\tr \vect{N}_l,\epsilon) +\frac{\beta}{K} \sum_{l=1}^K \vect{1}_S \tr \vect{N}_l N(\vect{\alpha}\tr \vect{N}_l,\epsilon)}\label{eq:non_UK0}\\
  &=& \E{\sum_{l=1}^{K} \phi_\beta(\vect{N}_{l})}\label{eq:non_UK1}\\
&\geq& \E{\min_{\substack{\vect{n}_{1},\cdots,\vect{n}_{|\mathcal{U}|}\in\mathbb{Z}^S_+:\\\sum_{l=1}^{|\mathcal{U}|} n_{l,s}=|\mathcal{U}^{(s)}|}} \sum_{l=1}^{|\mathcal{U}|} \phi_\beta(\vect{n}_{l})}\label{eq:non_UK2}
  \end{IEEEeqnarray}
  where \eqref{eq:non_UK1} is by the definition of $\phi_\beta(\cdot)$ in \eqref{eq:non_phi_def}. In \eqref{eq:non_UK2}, the expectation is only with respect to the random variables $|\mathcal{U}^{(1)}|, \cdots,|\mathcal{U}^{(S)}|$ and $|\mathcal{U}|$.
  Next,  \eqref{eq:non_UK2} is lower-bounded by using the lower convex envelope of $\phi_{\beta}(\cdot)$ and its convexity:
\begin{IEEEeqnarray}{rCl}
\IEEEeqnarraymulticol{3}{l}{\E{T}+\beta \E{P_1}}\nonumber\\\quad
 &\geq& \E{\min_{\substack{\vect{n}_{1},\cdots,\vect{n}_{|\mathcal{U}|}\in\mathbb{Z}_+^S:\\\sum_{l=1}^{|\mathcal{U}|} n_{l,s}=|\mathcal{U}^{(s)}|}} \sum_{l=1}^{|\mathcal{U}|} \breve{\phi}_\beta(\vect{n}_i)}\label{eq:non_UK5}\\
  &\geq& \E{\min_{\substack{\vect{n}_{1},\cdots,\vect{n}_{|\mathcal{U}|}\in\mathbb{Z}_+^S:\\\sum_{l=1}^{|\mathcal{U}|} n_{l,s}=|\mathcal{U}^{(s)}|}} |\mathcal{U}| \breve{\phi}_\beta\mathopen{}\Bigg(\frac{1}{|\mathcal{U}|}\sum_{l=1}^{|\mathcal{U}|}\vect{n}_{l}\Bigg)}\label{eq:non_UK6}\\ 
 &=& \E{|\mathcal{U}|\breve{\phi}_{\beta}\farg{ \left[\frac{|\mathcal{U}^{(1)}|}{|\mathcal{U}|},\cdots,\frac{|\mathcal{U}^{(S)}|}{|\mathcal{U}|}\right] }}.\label{eq:non_UK7}
\end{IEEEeqnarray}    
Here, \eqref{eq:non_UK5} follows because the lower convex envelope $\breve{\phi}_{\beta}(\cdot)$ of $\phi_{\beta}(\cdot)$ is smaller than or equal $\phi_{\beta}(\cdot)$ and \eqref{eq:non_UK6} follows from convexity of $\breve{\phi}_{\beta}(\cdot)$. The result follows by noting that the random vector $\left[|\mathcal{U}^{(1)}|,\cdots,|\mathcal{U}^{(S)}|,|\mathcal{U}^{(0)}|\right]$ is multinomial distributed with $S+1$ categories, $K$ trials, and event probabilities $[(1-q)p_1, \cdots, (1-q)p_S,q]$. 
\end{IEEEproof}
The following lemma shows that we can use the concavity of $N(\cdot,\epsilon)$ to simplify the computation of $\breve{\phi}_\beta(\cdot)$.
\begin{lemma}
\label{lem:breve_phi}
  For every $\beta > 0$, we have
  \begin{IEEEeqnarray}{rCl}
    \breve{\phi}_\beta(\vect{x}) = \min_{\substack{\vect{\zeta}\in\mathbb{R}^S:\\ \vect{1}_S\tr \vect{\zeta} = 1 \\\forall s: \zeta_s >0 }} \sum_{s=1}^S \zeta_s \breve{\phi}^{(s)}_\beta(x_s/\zeta_s)\label{eq:breve_simple_comp_claim}
  \end{IEEEeqnarray}
  where we have defined
  \begin{IEEEeqnarray}{rCl}
  \phi_\beta^{(s)}(x) \triangleq N(\alpha_s x,\epsilon)(1+\beta x/K)\label{eq:phis_def}
  \end{IEEEeqnarray}
  for $x\geq 0$ and $s\in\{1,\cdots,S\}$.
  Additionally, the optimization problem in \eqref{eq:breve_simple_comp_claim} is convex.
  \end{lemma}
\begin{IEEEproof}
See Appendix~\ref{app:lem_proof}
\end{IEEEproof}

For the case with fixed message sizes, i.e., when $S=1$, Proposition~\ref{prop:prop_converse} reduces to the following corollary.
\begin{corollary}
For every $\beta\geq 0$, we have
  \begin{IEEEeqnarray}{rCl} 
    \E{T} + \beta \E{P_1} \geq (1-q) K \breve{\phi}^{(1)}_{\beta}(1)
  \end{IEEEeqnarray}
  where $\phi^{(1)}_\beta(\cdot)$ is defined in \eqref{eq:phis_def}.
\end{corollary}
This readily follows from $\vect{L}_{1:S}/(\vect{1}_S\tr \vect{L}_{1:S}) = 1$ and because $\vect{1}_S\tr \vect{L}_{1:S}$ is Binomial distributed with parameters $K$ and $1-q$, and hence $\vect{1}_S\tr \vect{L}_{1:S} = (1-q)K$.

\subsection{Genie-aided protocol}\label{sec:genie}
We put forth a genie-aided protocol that uses the intuition obtained through Proposition~\ref{prop:prop_converse} and Lemma~\ref{lem:breve_phi}. Here, ``genie-aided'' refers to the fact that the protocol assumes that the knowledge about $\{D_k\}$ is available at all users. Lemma~2 suggests that one should group messages of the same sizes together rather than grouping messages of mixed message sizes. The purpose of introducing a genie-aided protocol is to show that it achieves a trade-off close to that of the lower bound. Moreover, we can compare the non-genie-aided protocols, introduced in Section~\ref{sec:protocol_design}, to the genie-aided protocol to show the impact of control information. Such comparisons are provided in Section~\ref{sec:numerical}. 

First, for a set of users $\mathcal{\bar U}\subseteq \mathcal{K}$, $V\in\mathbb{N}$, and $\varepsilon\in(0,1)$, we define a $(\mathcal{\bar U}, V, \varepsilon)$-protocol as follows. The users $\mathcal{\bar U}$ are divided into $G\triangleq\lceil |\mathcal{\bar U}|/V \rceil$ disjoint sets $\{\mathcal{\bar U}_l\}_{l\in\{1,\cdots,G\}}$ such that 
\begin{IEEEeqnarray}{rCl}
|\mathcal{\bar U}_l|&\triangleq& \left\{\begin{array}{ll}
\lfloor |\mathcal{\bar U}|/ G \rfloor+1, & l\in\{1,\cdots,\text{mod}(|\mathcal{\bar U}|, G)\}\\
\lfloor |\mathcal{\bar U}|/ G \rfloor, & \text{otherwise}.
\end{array}\right. 
\end{IEEEeqnarray}
One can verify that $\sum_{l=1}^G |\mathcal{\bar U}_l| = |\mathcal{\bar U}|$.
  Sequentially, for  $l\in\{1,\cdots,G\}$, the transmitter encodes and conveys a packet containing $\bigoplus_{k\in \mathcal{\bar U}_l}\mathrm{M}_k$ with error probability $\varepsilon$ using $N(\sum_{k\in\mathcal{\bar U}_l} D_k,\varepsilon)$ channel uses. Here, $\oplus$ denotes the concatenation of messages. While the number of channel uses spend at the transmitter is given  by $\sum_{i=1}^{G} N\farg{\sum_{k\in\mathcal{\bar U}_i} D_k,\epsilon}$, each user only needs to receive and decode one of the $G$ packets. We also note that a $(\mathcal{\bar U}, V, \varepsilon)$-protocol assumes control information at all the users $\mathcal{\bar U}$, i.e., the users needs to know $\mathcal{\bar U}$, $\{D_k\}_{k\in\mathcal{\bar U}}$, $V$, and $\varepsilon$.

For our genie-aided protocol, we define $\mathcal{U}^{(s)} \triangleq \{k\in \mathcal{K}: D_k = \alpha_s\}$, for $s\in\{0,\cdots, S\}$, and fix a vector $\vect{V}\in\mathcal{K}^S$. Now, sequentially for each $s\in\{1,\cdots,S\}$, the transmitter delivers the messages of the users $\mathcal{U}^{(s)}$ using a $(\mathcal{U}^{(s)}, V_s,\epsilon)$-protocol. We denote the average frame duration $\E{T}$ and the average power $\E{P_1}$ by $\bar T_{\text{genie}}^{(\vect{V},\epsilon)}$ and $\bar P_{\text{genie}}^{(\vect{V},\epsilon)}$, respectively. 

The vector $\vect{V}$ is left to be specified. We can trace of optimal trade-off between average frame duration and average power by solving the integer optimization problem for all $\beta \geq 0$:
\begin{IEEEeqnarray}{rCl}
  \min_{\vect{V} \in\mathcal{K}^S} \bar T_{\text{genie}}^{(\vect{V},\epsilon)} + \beta \bar P_{\text{genie}}^{(\vect{V},\epsilon)}
\end{IEEEeqnarray}

\section{Protocol Design}\label{sec:protocol_design}
In the following, we devise actual protocols that trade-off between average frame duration and average power consumption at the users. In contrast to the genie-aided protocol in Section~\ref{sec:genie}, these protocols need to convey control information.

\subsection{Fixed message size}
We initiate our discussion of protocol design with the case of fixed message size, i.e., $S=1$. In this case, the control information only consists  of which users are active. 
We divide the set of users $\mathcal{K}$ into $B \triangleq \lceil K/W\rceil$ disjoint subsets  $\mathcal{K}_1,\cdots,\mathcal{K}_{B}$ such that $\bigcup_{i=1}^{B}\mathcal{K}_i = \mathcal{K}$ and such that  
\begin{IEEEeqnarray}{rCl}
 |\mathcal{K}_i| &=& \left\{\begin{array}{ll}
   \lfloor K/B\rfloor + 1, & i\in\{1,\cdots,\text{mod}(K, B)\} \\
\lfloor K/B\rfloor,    & \text{otherwise}.
 \end{array}
 \right. 
\end{IEEEeqnarray}
Here, $W\in\mathbb{N}$ is a protocol parameter to be set. 
The subsets $\{\mathcal{K}_i\}$ of $\mathcal{K}$ are termed user groups (UG). The transmitter forms a packet that contains only the number of active users in each UG, i.e., the packet encodes the vector $\left[|\mathcal{U}\cap \mathcal{K}_1|, |\mathcal{U}\cap \mathcal{K}_2|,\cdots,|\mathcal{U}\cap \mathcal{K}_{B}|\right]$. This vector constitutes a first layer of control information and can be uniquely represented by at most $k_1 = \lceil\lceil K/W\rceil\log_2 W\rceil$ bits. We encode the control information by an optimal channel code with error probability not exceeding $\epsilon_1\in(0,1)$ which can be achieved by approximately $N(k_1,\epsilon_1)$ channel uses. 
 
After successfully decoding the first packet, the users know the number of users in each UG, and thereby the structure of the remaining part of the transmission. The second layer encodes control information and messages associated with each UG. Specifically, for the $i$-th UG, the transmitter needs to inform the users of the $i$-th UG about which $|\mathcal{U}\cap \mathcal{K}_i|$ users of $\mathcal{K}_i$ are active. Hence, the control information for the $i$-th UG, can be represented by $k_{2,i} \triangleq\Big\lceil\log_2 {{|\mathcal{K}_i|} \choose {|\mathcal{U}\cap \mathcal{K}_i|}}\Big\rceil$ bits and is conveyed by using an optimal code with error probability not exceeding $\epsilon_2\in(0,1)$, which requires approximately $N(k_{2,i},\epsilon_2)$ channel uses. Now, the messages of the active users in the $i$-th UG $\mathcal{U}_i \triangleq \mathcal{U}\cap \mathcal{K}_i$ are conveyed with error probability not exceeding $\epsilon_3\in(0,1)$ using an $(\mathcal{U}_i, V, \epsilon_3)$-protocol, where $V\in\mathcal{K}$ is another protocol parameter to be set. We emphasize that we can use an $(\mathcal{U}_i, V, \epsilon_3)$-protocol because the set of active users $\mathcal{U}_i$ knows $\mathcal{U}_i$ from the the control information provided that the first two packets are successfully decoded. Based on the description of the protocol above, one can compute $\E{T}$ and $\E{P_1}$ which we denote by $\bar T_{\text{fixed}}^{(V,W,\vect{\epsilon})}$ and $\bar P_{\text{fixed}}^{(V,W,\vect{\epsilon})}$, respectively. Here, $\vect{\epsilon}$ is the vector $[\epsilon_1,\epsilon_2,\epsilon_3]$.

The parameters $V$, $W$, and $\vect{\epsilon}$ are left to be specified. We can trace  the optimal achievable trade-off of the proposed protocol by solving the following optimization problem for all $\beta \geq 0$:
\begin{IEEEeqnarray}{rCl}
  \min_{(V,W)\in\mathcal{K}^2} \min_{\substack{\vect{\epsilon}\in[0,1]^3:\\\prod_{k=1}^3 (1-\epsilon_k) \geq 1-\epsilon}} \bar T_{\text{fixed}}^{(V,W,\vect{\epsilon})} + \beta\bar P_{\text{fixed}}^{(V,W,\vect{\epsilon})}.\label{eq:general_prot_opt}
\end{IEEEeqnarray}
While the outer minimization is an integer optimization problem which can only be solved using exhaustive search, the inner minimization is convex and can be solved using standard convex optimization algorithms. This is shown in the following lemma.
\begin{lemma}
  The inner optimization problem in \eqref{eq:general_prot_opt} is convex in $\vect{\epsilon}$. 
\end{lemma}
\begin{IEEEproof}
Note that, for fixed $V$ and $W$, the objective function in \eqref{eq:general_prot_opt} depends only on $\vect{\epsilon}$ through a nonnegative linear combination of $Q$-functions of $\epsilon_1$, $\epsilon_2$, and $\epsilon_3$, i.e., there exist nonnegative constants $a_1, a_2,$ and $a_3$ such that
\begin{IEEEeqnarray}{rCl}
\IEEEeqnarraymulticol{3}{l}{\bar T_{\text{fixed}}^{(V,W,\vect{\epsilon})} + \beta\bar P_{\text{fixed}}^{(V,W,\vect{\epsilon})}} \nonumber\\
&= a_{1} Q^{-1}(\epsilon_1) +a_{2} Q^{-1}(\epsilon_2) + a_{3} Q^{-1}(\epsilon_3). \label{eq:obj_func}
\end{IEEEeqnarray}
This is because $\bar T_{\text{fixed}}^{(V,W,\vect{\epsilon})}$ and $\beta\bar P_{\text{fixed}}^{(V,W,\vect{\epsilon})}$ are evaluated using $N(k,\varepsilon)$.
To show convexity of the optimization problem \eqref{eq:general_prot_opt}, we use the substitution $\epsilon_i = 1-\ee{u_i}$ for $u_i \leq 0$ and $i\in\{1,2,3\}$, which yields the equivalent constraint $u_1+u_2+u_3=\log\left(\prod_{k=1}^3 (1-\epsilon_k)\right)  \leq \log(1-\epsilon)$ which is linear. Consequently, it is sufficient to show that $Q^{-1}(1-\ee{u_i})$ is convex for $i\in\{1,2,3\}$. This follows because the logarithm of the cumulative distribution function of the Gaussian distribution $f(x)\triangleq \log(1-Q(x))$ is concave and increasing. Thus, its inverse function $f^{-1}(x) = Q^{-1}(1-\ee{x})$ is convex and increasing. 
\end{IEEEproof}

At this point, we have not discussed the possibility of undetected errors. Approximations like \eqref{eq:fbl} do not give any guarentee for the probability of detecting an error. Using CRCs, the probability of undetected error can be made arbitrarily small, but it is always positive and less than or equal $\epsilon$. Suppose that decoding of the first packet, containing control information, fails for the $k$-th user. In this case, the subsequent behavior is random, and the $k$-th user will (with high probability) not correctly decode the following packets. However, since the packet sizes are limited by $\alpha_S$, we can compute the worst-case power consumption at the users, say $P_{\text{worst}}$. We then cope with the problem of undetected errors simply by adding, to the power consumption at each user, the term $\epsilon P_{\text{worst}}$, which corresponds to the worst-case contribution to the power consumption. 
 
\subsection{Variable message size}
\label{sec:varying} 
\begin{figure}[!t]
  \centering
  \includegraphics[width=3.4in]{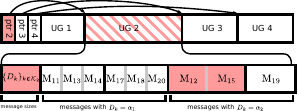}
  \caption{An example of the protocol in Section~\ref{sec:varying} with $S= 2$, $K=40$, $W=10$, and $\vect{V} = [3,2]$. Packets surrounded by black separators corresponds to an encoded packet. Grey separators means ``encoded jointly'', e.g., the messages $\mathrm{M}_{12}$ and $\mathrm{M}_{15}$ are jointly encoded in one packet. The red shaded parts of the protocol depicts the packets that the users $12$ and $15$ needs to decode.}
  \label{fig:protocol}
\end{figure}
Next, we consider the case $S \geq 2$. 
The users are grouped into $B\triangleq \lceil K/W\rceil$ UGs in the same way as for the fixed message size protocol. The UGs are encoded sequentially after the control information of the first layer. The control information of the first layer consists of pointers to the time indices of the beginning of each UG. Thus, based on the control information of the first layer, each user can identify the location of its UG. Note that we need only $B-1$ pointer because the first UG is transmitted immediately after the control information.  Each pointer is encoded separately in a packet using an optimal code with an error probability not exceeding $\epsilon_1$. Observe that one can compute the maximum length (in channel uses) of each UG and thereby the number of bits required for each pointer.

The control information of the second layer for the $i$-th UG consists of $\{D_k\}_{k\in\mathcal{K}_i}$, represented by $\lceil |\mathcal{K}_i| \log_2 (S+1)  \rceil$ bits. These bits are transmitted using an optimal code with error probability not exceeding $\epsilon_2$.
\begin{comment}
of a vector with the number of users in $\mathcal{K}_i$ having $D_k=0$, $D_k=\alpha_1, \cdots, D_k = \alpha_S$, i.e.,
\begin{multline}
 \vect{a}\triangleq \big[  \mathcal{K}_i\cap \{k: D_k = 0\}, \mathcal{K}_i\cap \{k: D_k = \alpha_1\}\\,\cdots,\mathcal{K}_i\cap \{k: D_k = \alpha_S\}  \big].\label{eq:vecta}
\end{multline}
Here, we denote the entries of $\vect{a}$ by $\vect{a}_0,\cdots,\vect{a}_S$.
This vector is multinomial distributed with $S+1$ categories, $|\mathcal{K}_i|$ trials, and event probabilities $[q, (1-q)p_1, \cdots,(1-q)p_S]$. This vector can take $\multiset{|\mathcal{K}_i|}{S+1}$ distinct values, where $\multiset{n}{k}$ denotes the number of multisets of cardinality $k$ that can be drawn from a set of cardinality $n$. This vector is conveyed using an optimal code with error probability not exceeding $\epsilon_2$. 
\end{comment}
Finally, sequentially for each $s\in\{1,\cdots,S\}$, the transmitter encodes the messages of the users $\mathcal{U}_i^{(s)}$ using an $(\mathcal{U}_i^{(s)},V_s,\epsilon_3)$-protocol, where $\vect{V} = [V_1,\cdots,V_S]$ are protocol parameters to be specified. The protocol is illustrated in Fig.~\ref{fig:protocol}.

We denote $\E{T}$ and $\E{P_1}$ by $\bar T_{\text{variable}}^{(\vect{V},W,\vect{\epsilon})}$ and $\bar P_{\text{variable}}^{(\vect{V},W,\vect{\epsilon})}$, respectively, and optimize the parameters of the protocol using the optimization problem
\begin{IEEEeqnarray}{rCl}
  \min_{(\vect{V},W)\in\mathcal{K}^{S+1}} \min_{\substack{\vect{\epsilon}\in[0,1]^3:\\ \prod_{k=1}^3 (1-\epsilon_k) \geq 1- \epsilon} } \bar T_{\text{variable}}^{(\vect{V}, W,\vect{\epsilon})} +\beta \bar P_{\text{variable}}^{(\vect{V}, W,\vect{\epsilon})}.
\end{IEEEeqnarray}  
As for the fixed message size protocol, the inner minimization is convex.

\section{Numerical Results}\label{sec:numerical} 
In this section, we plot the lower bound along with the optimal achievable trade-offs for the proposed protocols. All results are for $\epsilon=10^{-4}$, $P=0\ \text{dB}$, $q=0.5$.

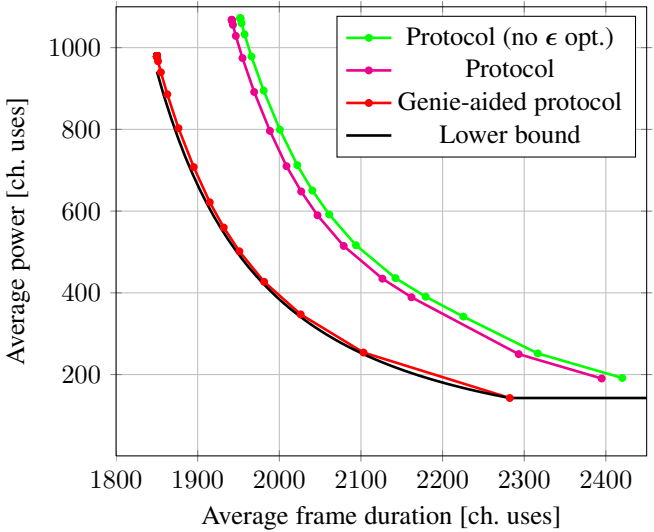
\begin{figure}[!t]
\centering
\begin{tikzpicture}
\begin{axis}[
/pgf/number format/.cd, 1000 sep={}, 
grid=both,
tick align=center,
width=0.4\paperwidth,
height=0.35\paperwidth,
ymin=1,
xmin=1800, 
ymax=1100,
xmax=2450, 
xlabel={Average frame duration [ch. uses]},
ylabel={Average power [ch. uses]},
legend entries = { {}{Protocol (no $\vect{\epsilon}$ opt.)}, {}{Protocol}, {}{Genie-aided protocol}, {}{Lower bound}}
] 
\addplot [
    color=green,
    mark=*,
    line width=1.0pt,
    mark size=1pt] table [x index = {0}, y index = {1}, col sep=comma]{fixed_plot_K16_P1.0_q0.50_alpha100_fixedconcat.dat};
\addplot [  
    color=magenta,   
    mark=*,  
    line width=1.0pt,
    mark size=1pt] table [x index = {0}, y index = {1}, col sep=comma]{fixed_plot_K16_P1.0_q0.50_alpha100_fixedopt.dat};
    \addplot [  
    color=red, mark=*, line width=1.0pt,mark size=1pt] table [x index = {0}, y index = {1}, col sep=comma]{fixed_plot_K16_P1.0_q0.50_alpha100_genie.dat};
        \addplot [line width=1.0pt,color=black, solid] table [x index = {0}, y index = {1}, col sep=comma]{fixed_plot_K16_P1.0_q0.50_alpha100_lower.dat};
\end{axis}
\end{tikzpicture}
\caption{Trade-off between average transmission time and average power consumption for the case $K=16$, $P=1$, $q=0.5$, $\alpha = 100$, $S=1$, and $\epsilon=10^{-4}$. Here, ``Protocol'' refers to the fixed message size protocol, while ``Protocol (no $\vect{\epsilon}$ opt.)'' refers to the fixed message size protocol with $\epsilon_1=\epsilon_2 = \epsilon_3 = 1-(1-\epsilon)^{1/3})$.}
\label{fig:trade_off_16_100}
\end{figure}

\begin{figure}[!t]
\centering
\begin{tikzpicture}
\begin{axis}[
/pgf/number format/.cd, 1000 sep={},
grid=both,
tick align=center,
width=0.4\paperwidth,
height=0.35\paperwidth,
ymin=0,
xmin=130000, 
ymax=35000,
xmax=150000, 
xlabel={Average frame duration [ch. uses]},
ylabel={Average power [ch. uses]},
legend entries = { {}{Protocol (no $\vect{\epsilon}$ opt.)}, {}{Protocol}, {}{Genie-aided protocol}, {}{Lower bound}}
] 
\addplot [
    color=green,
    mark=*,
    line width=1.0pt,
    mark size=1pt] table [x index = {0}, y index = {1}, col sep=comma]{fixed_plot_K128_P1.0_q0.50_alpha1000_fixedconcat.dat};
\addplot [  
    color=magenta,   
    mark=*,  
    line width=1.0pt,
    mark size=1pt] table [x index = {0}, y index = {1}, col sep=comma]{fixed_plot_K128_P1.0_q0.50_alpha1000_fixedopt.dat};
    \addplot [  
    color=red, mark=*, line width=1.0pt,mark size=1pt] table [x index = {0}, y index = {1}, col sep=comma]{fixed_plot_K128_P1.0_q0.50_alpha1000_genie.dat};
        \addplot [line width=1.0pt,color=black, solid] table [x index = {0}, y index = {1}, col sep=comma]{fixed_plot_K128_P1.0_q0.50_alpha1000_lower.dat};
\end{axis}
\end{tikzpicture}
\caption{Trade-off between average transmission time and average power consumption for the case $K=128$, $P=1$, $q=0.5$, $\alpha = 1000$, $S=1$, and $\epsilon=10^{-4}$. Here, ``Protocol'' refers to the fixed message size protocol, while ``Protocol (no $\vect{\epsilon}$ opt.)'' refers to the fixed message size protocol with $\epsilon_1=\epsilon_2 = \epsilon_3 = 1-(1-\epsilon)^{1/3})$.} 
\label{fig:trade_off_128_1000}
\end{figure}
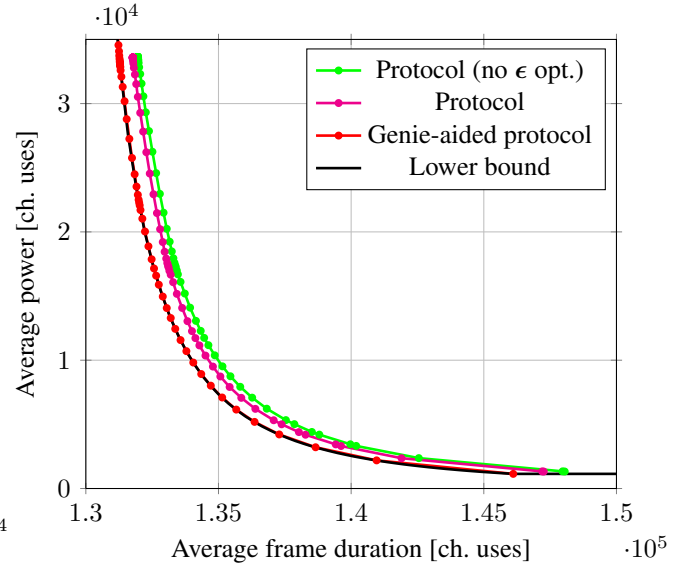

We first present results for the case with fixed message size. Fig.~\ref{fig:trade_off_16_100} and Fig.~\ref{fig:trade_off_128_100} show the trade-offs for $\alpha_1=100$ and $S=1$ for $K=16$ and $K=128$, respectively. We plot the lower bound given by Proposition~\ref{prop:prop_converse} and the trade-offs achievable by the genie-aided protocol and the fixed message size protocol. For the fixed message size protocol, we also plot the trade-off for the case where the inner minimization in \eqref{eq:general_prot_opt} is not performed and $\epsilon_1, \epsilon_2, \epsilon_3$ are set equally to $1-(1-\epsilon)^{1/3}$. For the protocols, we plot the lower convex envelopes and note that any point on them can be achieved by time-sharing between two sets of protocol parameters. We observe, as expected, that differences between the genie-aided protocols and the lower bounds are negligible. Optimizing over $\vect{\epsilon}$ also improves the trade-off slightly. This happens because the control information which is destined to many users needs better protection compared to a group of messages destined only to a group of users. Finally, we observe a significant gap between the genie-aided protocol and the fixed message size protocol which reflects the significance of control information for broadcast of small messages. Fig.~\ref{fig:trade_off_16_1000} and Fig.~\ref{fig:trade_off_128_1000} shows the trade-offs for $\alpha=1000$. In this case, we see that the gap between the genie-aided protocol and the fixed message size protocol becomes less significant.

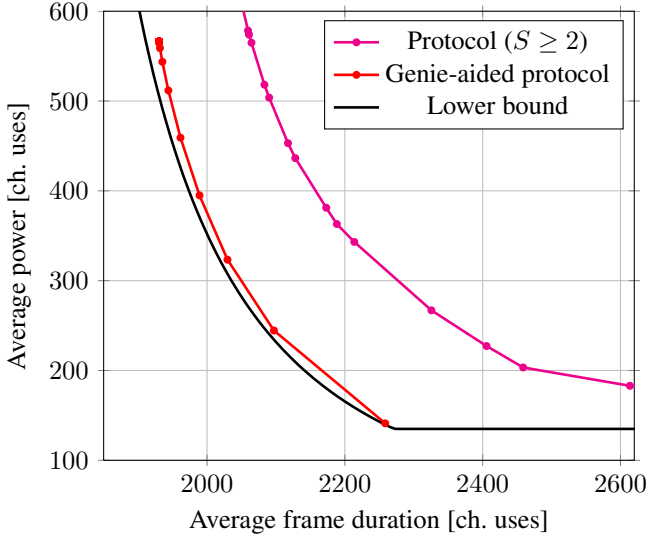
\begin{figure}[!t]
\centering
\begin{tikzpicture}
\begin{axis}[
/pgf/number format/.cd, 1000 sep={},
grid=both,
tick align=center,
width=0.4\paperwidth,
height=0.35\paperwidth,
ymin=100,
xmin=1850, 
ymax=600,
xmax=2620, 
xlabel={Average frame duration [ch. uses]},
ylabel={Average power [ch. uses]},
legend entries = { {}{Protocol ($S\geq 2$)}, {}{Genie-aided protocol}, {}{Lower bound}}
] 
\addplot [color=magenta, mark=*, line width=1.0pt, mark size=1pt] 
    table [x index = {0}, y index = {1}, col sep=comma]{two_mess_plot_K16_P1.0_q0.50_alpha50_150_opt.dat};
\addplot [color=red, mark=*, line width=1.0pt,mark size=1pt] 
  table [x index = {0}, y index = {1}, col sep=comma]{two_mess_plot_K16_P1.0_q0.50_alpha50_150_genie.dat};
\addplot [line width=1.0pt,color=black, solid] 
  table [x index = {0}, y index = {1}, col sep=comma]{two_mess_plot_K16_P1.0_q0.50_alpha50_150_lower.dat};
\end{axis}
\end{tikzpicture}
\caption{Trade-off between average transmission time and average power consumption for the case $K=16$, $P=1$, $q=0.5$, $\vect{p}=[0.5,0.5]$, $\vect{\alpha} = [50,150]$, $S=2$, and $\epsilon=10^{-4}$.}
\label{fig:trade_off_two_mess_16_100}
\end{figure}

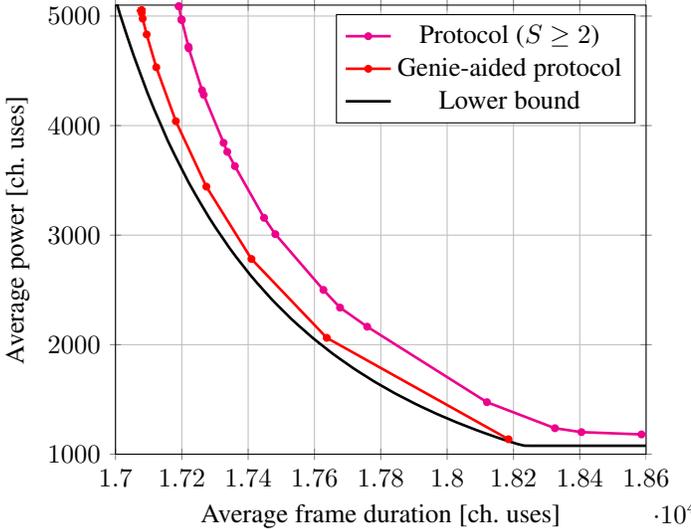
\begin{figure}[!t]
\centering
\begin{tikzpicture}
\begin{axis}[
/pgf/number format/.cd, 1000 sep={},
grid=both,
tick align=center,
width=0.4\paperwidth,
height=0.35\paperwidth,
ymin=1000,
xmin=17000, 
ymax=5100,
xmax=18600, 
xlabel={Average frame duration [ch. uses]}, 
ylabel={Average power [ch. uses]},
legend entries = { {}{Protocol ($S\geq 2$)}, {}{Genie-aided protocol}, {}{Lower bound}}
] 
\addplot [color=magenta, mark=*, line width=1.0pt, mark size=1pt] 
    table [x index = {0}, y index = {1}, col sep=comma]{two_mess_plot_K16_P1.0_q0.50_alpha500_1500_opt.dat};
\addplot [color=red, mark=*, line width=1.0pt,mark size=1pt] 
  table [x index = {0}, y index = {1}, col sep=comma]{two_mess_plot_K16_P1.0_q0.50_alpha500_1500_genie.dat};
\addplot [line width=1.0pt,color=black, solid] 
  table [x index = {0}, y index = {1}, col sep=comma]{two_mess_plot_K16_P1.0_q0.50_alpha500_1500_lower.dat};
\end{axis}
\end{tikzpicture}
\caption{Trade-off between average transmission time and average power consumption for the case $K=16$, $P=1$, $q=0.5$, $\vect{p}=[0.5,0.5]$, $\vect{\alpha} = [500,1500]$, $S=2$, and $\epsilon=10^{-4}$.}
\label{fig:trade_off_two_mess_16_1000}
\end{figure}

Finally, in Fig.~\ref{fig:trade_off_two_mess_16_100} and Fig.~\ref{fig:trade_off_two_mess_16_1000}, we depict the trade-offs for $K=16$, $\vect{p}=[0.5,0.5]$ and with $\vect{\alpha} = [50,150]$ and $\vect{\alpha}=[500,1500]$, respectively. Our observations are similar to those for the fixed message size protocol.

\begin{figure}[!t]
\centering
\begin{tikzpicture}
\begin{axis}[
/pgf/number format/.cd, 1000 sep={},
grid=both,
tick align=center,
width=0.4\paperwidth,
height=0.35\paperwidth,
ymin=0,
xmin=13500, 
ymax=4000,
xmax=19500, 
xlabel={Average frame duration [ch. uses]},
ylabel={Average power [ch. uses]},
legend entries = { {}{Protocol (no $\vect{\epsilon}$ opt.)}, {}{Protocol}, {}{Genie-aided protocol}, {}{Lower bound}}
] 
\addplot [
    color=green,
    mark=*,
    line width=1.0pt,
    mark size=1pt] table [x index = {0}, y index = {1}, col sep=comma]{fixed_plot_K128_P1.0_q0.50_alpha100_fixedconcat.dat};
\addplot [  
    color=magenta,   
    mark=*,  
    line width=1.0pt,
    mark size=1pt] table [x index = {0}, y index = {1}, col sep=comma]{fixed_plot_K128_P1.0_q0.50_alpha100_fixedopt.dat};
    \addplot [  
    color=red, mark=*, line width=1.0pt,mark size=1pt] table [x index = {0}, y index = {1}, col sep=comma]{fixed_plot_K128_P1.0_q0.50_alpha100_genie.dat};
        \addplot [line width=1.0pt,color=black, solid] table [x index = {0}, y index = {1}, col sep=comma]{fixed_plot_K128_P1.0_q0.50_alpha100_lower.dat};
\end{axis}
\end{tikzpicture}
\caption{Trade-off between average transmission time and average power consumption for the case $K=128$, $P=1$, $q=0.5$, $\alpha = 100$, $S=1$, and $\epsilon=10^{-4}$. Here, ``Protocol'' refers to the fixed message size protocol, while ``Protocol (no $\vect{\epsilon}$ opt.)'' refers to the fixed message size protocol with $\epsilon_1=\epsilon_2 = \epsilon_3 = 1-(1-\epsilon)^{1/3})$.} 
\label{fig:trade_off_128_100}
\end{figure}

\begin{figure}[!t]
\centering
\begin{tikzpicture}
\begin{axis}[
/pgf/number format/.cd, 1000 sep={},
grid=both,
tick align=center,
width=0.4\paperwidth,
height=0.35\paperwidth,
ymin=0,
xmin=16700, 
ymax=9000,
xmax=18500, 
xlabel={Average frame duration [ch. uses]},
ylabel={Average power [ch. uses]},
legend entries = { {}{Protocol (no $\vect{\epsilon}$ opt.)}, {}{Protocol}, {}{Genie-aided protocol}, {}{Lower bound}}
] 
\addplot [
    color=green,
    mark=*,
    line width=1.0pt,
    mark size=1pt] table [x index = {0}, y index = {1}, col sep=comma]{fixed_plot_K16_P1.0_q0.50_alpha1000_fixedconcat.dat};
\addplot [  
    color=magenta,   
    mark=*,  
    line width=1.0pt,
    mark size=1pt] table [x index = {0}, y index = {1}, col sep=comma]{fixed_plot_K16_P1.0_q0.50_alpha1000_fixedopt.dat};
    \addplot [  
    color=red, mark=*, line width=1.0pt,mark size=1pt] table [x index = {0}, y index = {1}, col sep=comma]{fixed_plot_K16_P1.0_q0.50_alpha1000_genie.dat};
        \addplot [line width=1.0pt,color=black, solid] table [x index = {0}, y index = {1}, col sep=comma]{fixed_plot_K16_P1.0_q0.50_alpha1000_lower.dat};
\end{axis}
\end{tikzpicture}
\caption{Trade-off between average transmission time and average power consumption for the case $K=16$, $P=1$, $q=0.5$, $\alpha = 1000$, $S=1$, and $\epsilon=10^{-4}$. Here, ``Protocol'' refers to the fixed message size protocol, while ``Protocol (no $\vect{\epsilon}$ opt.)'' refers to the fixed message size protocol with $\epsilon_1=\epsilon_2 = \epsilon_3 = 1-(1-\epsilon)^{1/3})$.}
\label{fig:trade_off_16_1000}
\end{figure}

\section{Conclusion}\label{sec:conclusions}
In this paper, we considered the AWGN broadcast channel with $K$ users with symmetric channel conditions. The downlink transmission is organized in frames. In each frame, a message of random size (in bits) is destined to each of the users in such a way that the message sizes are unknown to the users. The message can also be of size zero, which means the user should not receive data in that frame. A user, however, still needs to decode a certain amount of information from the frame in order to learn that there is no data destined to her in this particular frame. Hence, in addition to the messages, a protocol needs to convey control information that describes the structure of the transmission and the sizes of the messages. We used approximations of the maximum coding rate for the AWGN channel from finite blocklength information theory to show that jointly encoding different groupings of the messages enable the protocol designer to trade-off between average frame duration and the average power consumption at the users. Specifically, we derived a lower bound for the trade-off curve which assumed that control information was available at the users, a genie-aided protocol, and two practical protocols. Our numerical results showed that the genie-aided protocol achieved a trade-off curve that closely matched the lower bound. For both of our practical protocols, the control information led to a significantly worse trade-off curves when the messages were small and when compared to the genie-aided protocol.  There are several directions for future research: 
\begin{enumerate} 
    \item In Section~\ref{sec:system_model}, we significantly restricted our general system model to a space of practical and tractable protocols. A rigorous information-theoretic treatment of our general system model might lead to improved protocols and lower bounds. 
    \item The system model has two obvious extensions: one can extend the system model to include fading, and one can introduce asymmetric channel conditions using results from \cite{Yang2014}.     
    \item While we are able to quantify the suboptimality of our protocols by comparison to the lower bound, our protocols are still heuristic. One interesting idea for future research is to systematically investigate the design of good protocols that include control information.
\end{enumerate}

\appendices 

\section{Proof of Lemma~\ref{lem:breve_phi}}\label{app:lem_proof}
By definition of the lower convex envelope, for every $\vect{x}\in\mathbb{R}_+^S$, there exists a vector $\vect{\nu}\in\mathbb{R}_+^S$ with $\vect{1}_S\tr \vect{\nu}=1$ and $I$ points $\vect{a}_i\in\mathbb{R}_+^S$, for $i\in\{1,\cdots,I\}$, such that 
  \begin{IEEEeqnarray}{rCl}
    \breve{\phi}_{\beta}(\vect{x}) = \sum_{i=1}^{I} \nu_i \phi_\beta(\vect{a}_i)\label{eq:remark_convex_combination}
  \end{IEEEeqnarray}
  and such that
  \begin{IEEEeqnarray}{rCl}
    \vect{x} = \sum_{i=1}^I \nu_i \vect{a}_i.\label{eq:sum_to_x}
  \end{IEEEeqnarray}
Since $N(n,\epsilon)$ is concave in $n$, we have that $\phi_\beta(\vect{x})$ is concave on the simplex $\mathcal{A}_{\kappa}\triangleq \{\vect{x}\in\mathbb{R}^S_+: \vect{1}_{S}\tr \vect{x} = \kappa\}$ for every $\kappa\in\mathbb{R}_+$.  Consequently, for $i\in\{1,\cdots,I\}$, we have
\begin{IEEEeqnarray}{rCl}
\phi_\beta(\vect{a}_i) &=&\phi_\beta\farg{ \sum_{s=1}^S \frac{a_{i,s}}{\vect{1}_S\tr \vect{a}_i} \vect{\bar 0}^S_s\mathopen{}\left(\vect{1}_S\tr \vect{a}_i \right)} \label{eq:ineqm1}\\
 &\geq& \sum_{s=1}^S \frac{a_{i,s}}{\vect{1}_S\tr \vect{a}_i}\phi_\beta\farg{ \vect{\bar 0}^S_s\mathopen{}\left(\vect{1}_S\tr \vect{a}_i\right)}\label{eq:ineq0}\\
&=&  \sum_{s=1}^S \frac{a_{i,s}}{\vect{1}_S\tr \vect{a}_i}\phi^{(s)}_\beta\farg{\vect{1}_S\tr \vect{a}_i }.\label{eq:ineq1}
\end{IEEEeqnarray}
In \eqref{eq:ineqm1}, $\vect{0}_s^S(x)$ denotes an $S$-dimensional vector with $x$ in the $s$-th entry and zeroes in the rest, \eqref{eq:ineq0} follows by Jensen's inequality (concave) applied to $\phi_\beta(\vect{x})$ on the simplex $\mathcal{A}_{\vect{1}_S\tr \vect{a}_i}$, and \eqref{eq:ineq1} is by the definition of $\phi_{\beta}^{(s)}(\cdot)$ in \eqref{eq:phis_def}.
bWe can now lower-bound \eqref{eq:remark_convex_combination} as
\begin{IEEEeqnarray}{rCl}
  \breve{\phi}_{\beta}(\vect{x}) &\geq& \sum_{s=1}^S \sum_{i=1}^{I}  \frac{\nu_i a_{i,s}}{\vect{1}_S\tr \vect{a}_i} \phi^{(s)}_\beta\farg{\vect{1}_S\tr \vect{a}_i }\label{eq:ineq2}\\
     &\geq& \sum_{s=1}^S \sum_{i=1}^{I}  \frac{\nu_i a_{i,s}}{\vect{1}_S\tr \vect{a}_i} \breve{\phi}^{(s)}_\beta\farg{\vect{1}_S\tr \vect{a}_i }\label{eq:ineq4}\\
  &\geq& \sum_{s=1}^S \left(\sum_{i=1}^{I}  \frac{\nu_i a_{i,s}}{\vect{1}_S\tr \vect{a}_i} \right) \breve{\phi}^{(s)}_\beta\farg{ \frac{\sum_{i=1}^I\frac{\nu_i a_{i,s}}{\vect{1}_S\tr \vect{a}_i} \vect{1}_S\tr \vect{a}_i }{\sum_{i=1}^I \frac{\nu_i a_{i,s}}{\vect{1}_S\tr \vect{a}_i}}  }\label{eq:ineq3}\\
        &=& \sum_{s=1}^S \zeta_s \breve{\phi}^{(s)}_\beta\farg{x_s/\zeta_s}.\label{eq:ineq5}
\end{IEEEeqnarray}
Here, \eqref{eq:ineq2} is by \eqref{eq:remark_convex_combination} and \eqref{eq:ineq1}, \eqref{eq:ineq4} is by $\phi^{(s)}_\beta(x) \geq \breve{\phi}^{(s)}_\beta(x)$ for $x\geq 0$, \eqref{eq:ineq3} follows by Jensen's inequality (convex) applied to $\breve{\phi}^{(s)}_\beta(\cdot)$, and \eqref{eq:ineq5} follows by setting $\zeta_s \triangleq \sum_{i=1}^{I}  \frac{\nu_i a_{i,s}}{\vect{1}_S\tr \vect{a}_i}$ and by using $\sum_{i=1}^I\nu_i a_{i,s} = x_s$ by \eqref{eq:sum_to_x}. Thus, we have shown that the LHS of \eqref{eq:breve_simple_comp_claim} is larger than or equal the RHS of \eqref{eq:breve_simple_comp_claim}. 

Next, we establish the equality in \eqref{eq:breve_simple_comp_claim}. Suppose, on the contrary, that there exists a positive vector $\vect{\bar \zeta}\in\mathbb{R}^S$ such that $\vect{1}_S\tr \vect{\zeta} = 1$ and such that $\breve{\phi}_\beta(\vect{x}) > \sum_{s=1}^S \bar\zeta_s \breve{\phi}_\beta^{(s)}(x_s/\bar \zeta_s)$. This implies a contradiction:
\begin{IEEEeqnarray}{rCl}
\breve{\phi}_\beta(\vect{x}) &>& \sum_{s=1}^S \bar\zeta_s \breve{\phi}_\beta^{(s)}(x_s/\bar \zeta_s) \\
&=& \sum_{s=1}^S \bar\zeta_s \breve{\phi}_\beta(\vect{\bar 0}_s^S(x_s/\zeta_s)) \\
&\geq& \breve{\phi}_\beta(\vect{x}).\label{eq:contradiction}
\end{IEEEeqnarray}
Here, \eqref{eq:contradiction} follows by Jensen's inequality (convex) applied to $\breve{\phi}_\beta(\cdot)$. We conclude that \eqref{eq:breve_simple_comp_claim} must be satisfied with equality. Note that it is sufficient to write minimum instead of infimum in \eqref{eq:breve_simple_comp_claim} because we have shown the existence of a feasible point in \eqref{eq:breve_simple_comp_claim} that attains the minimum.

To show convexity of the optimization problem in \eqref{eq:breve_simple_comp_claim}, it is sufficient to show that the function $x \breve{\phi}^{(s)}(y/x)$ is convex in $x>0$ for a constant $y>0$ and $s\in\{1,\cdots,S\}$, i.e., for every $x_1>x_2>0$ and $\alpha \in[0,1]$, we need to show that
\begin{IEEEeqnarray}{rCl}
   \IEEEeqnarraymulticol{3}{l}{\alpha x_1 \breve{\phi}^{(s)}(y/x_1) + (1-\alpha)x_2 \breve{\phi}^{(s)}(y/x_2)}\nonumber\\\quad
  &\geq& (\alpha x_1 + (1-\alpha)x_2 )\breve{\phi}^{(s)}( y/(\alpha x_1 + (1-\alpha)x_2) ).\label{eq:convexity}
\end{IEEEeqnarray}
Fix, without loss of generality, arbitrary $x_1 > x_2 > 0$, $\alpha\in[0,1]$, and $s\in\{1,\cdots,S\}$. 
  Define the function
\begin{IEEEeqnarray}{rCl}
  g(x) = \frac{\frac{y}{x_2} -x}{\frac{y}{x_2} - \frac{y}{x_1}}  \breve{\phi}^{(s)}\farg{\frac{y}{x_1}} +\frac{x - \frac{y}{x_1}}{\frac{y}{x_2} - \frac{y}{x_1}}  \breve{\phi}^{(s)}\farg{\frac{y}{x_2}}.\IEEEeqnarraynumspace
\end{IEEEeqnarray}
Note that $x g(y/x)$ and $g(x)$ are affine functions in $x>0$ and that $g(y/x_1) = \breve{\phi}^{(s)}(y/x_1)$ and $g(y/x_2) = \breve{\phi}^{(s)}(y/x_2)$. Thus, since $\breve{\phi}^{(s)}(\cdot)$ is convex, we have $g(x) \geq \breve{\phi}^{(s)}(x)$ for $x\in[y/x_1, y/x_2]$. To verify \eqref{eq:convexity}, we write 
\begin{IEEEeqnarray}{rCl}
  \IEEEeqnarraymulticol{3}{l}{\alpha x_1 \breve{\phi}^{(s)}(y/x_1) + (1-\alpha)x_2 \breve{\phi}^{(s)}(y/x_2)}\nonumber\\\quad &=& (\alpha x_1 + (1-\alpha)x_2)g( y/(\alpha x_1 + (1-\alpha)x_2) )\IEEEeqnarraynumspace\\
  &\geq& (\alpha x_1 + (1-\alpha)x_2)\breve{\phi}^{(s)}( y/(\alpha x_1 + (1-\alpha)x_2) )\label{eq:convexity_argu}
\end{IEEEeqnarray}
This establishes the convexity of the optimization problem in \eqref{eq:breve_simple_comp_claim} because we can redo the above argument for all $x_2>x_1>0$ and $\alpha\in[0,1]$.

%The objective is to minimize the frame duration 
%\begin{align}
  
%\end{align}

% or




\bibliographystyle{IEEEtranTCOM}
\bibliography{comb_header_and_data}

% Generated by IEEEtranTCOM.bst, version: 1.13 (2008/09/30)
\begin{thebibliography}{1}
\baselineskip 12pt
\providecommand{\url}[1]{#1}
\csname url@samestyle\endcsname
\providecommand{\newblock}{\relax}
\providecommand{\bibinfo}[2]{#2}
\providecommand{\BIBentrySTDinterwordspacing}{\spaceskip=0pt\relax}
\providecommand{\BIBentryALTinterwordstretchfactor}{4}
\providecommand{\BIBentryALTinterwordspacing}{\spaceskip=\fontdimen2\font plus
\BIBentryALTinterwordstretchfactor\fontdimen3\font minus
  \fontdimen4\font\relax}
\providecommand{\BIBforeignlanguage}[2]{{%
\expandafter\ifx\csname l@#1\endcsname\relax
\typeout{** WARNING: IEEEtran.bst: No hyphenation pattern has been}%
\typeout{** loaded for the language `#1'. Using the pattern for}%
\typeout{** the default language instead.}%
\else
\language=\csname l@#1\endcsname
\fi
#2}}
\providecommand{\BIBdecl}{\relax}
\BIBdecl

\bibitem{Popovski2014}
P.~Popovski, ``{Ultra-reliable communication in 5G wireless systems},'' in
  \emph{IEEE Int. Conf. 5G for Ubiquitous Connectivity}, Levi, Finland, Nov.
  2014, pp. 146--151.

\bibitem{Durisi2015}
G.~Durisi, T.~Koch, and P.~Popovski, ``Towards massive, ultra-reliable, and
  low-latency wireless: The art of sending short packets,'' \emph{Proc. IEEE},
  2016, to appear.

\bibitem{FiveDisruptive}
F.~Boccardi, R.~W. Heath, A.~Lozano, T.~L. Marzetta, and P.~Popovski, ``{Five
  disruptive technology directions for 5G},'' \emph{IEEE Commun. Mag.},
  vol.~52, no.~2, pp. 74--80, 2014.

\bibitem{Strassen2009}
V.~Strassen, ``{Asymptotische abschätzungen in Shannon's
  informationstheorie},'' in \emph{Trans. 3rd Prague Conf. Int. Theory},
  Prague, Czech Republic, 1962, pp. 689--723.

\bibitem{Polyanskiy2010b}
Y.~Polyanskiy, H.~V. Poor, and S.~Verd\'{u}, ``Channel coding rate in the
  finite blocklength regime,'' \emph{IEEE Trans. Inf. Theory}, vol.~56, no.~5,
  pp. 2307--2359, 2010.

\bibitem{Xu2016}
\BIBentryALTinterwordspacing
S.~Xu, T.-H. Chang, S.-C. Lin, C.~Shen, and G.~Zhu, ``Energy-efficient packet
  scheduling with finite blocklength codes: Convexity analysis and efficient
  algorithms,'' \emph{arXiv}, pp. 1--30, Mar. 2016. [Online]. Available:
  \url{http://arxiv.org/pdf/1603.03133.pdf}
\BIBentrySTDinterwordspacing

\bibitem{Makki2014}
B.~Makki, T.~Svensson, and M.~Zorzi, ``Finite block-length analysis of the
  incremental redundancy harq,'' \emph{IEEE Wireless Commun. Letters.}, vol.~3,
  no.~5, pp. 529--532, 2014.

\bibitem{Khalili2015}
\BIBentryALTinterwordspacing
S.~Khalili and O.~Simeone, ``Uplink harq for distributed and cloud ran via
  separation of control and data planes,'' \emph{arXiv}, pp. 1--27, Dec. 2015.
  [Online]. Available: \url{http://arxiv.org/pdf/1508.06570v3.pdf}
\BIBentrySTDinterwordspacing

\bibitem{Yang2014}
W.~Yang, G.~Durisi, T.~Koch, and Y.~Polyanskiy, ``{Quasi-static
  multiple-antenna fading channels at finite blocklength},'' \emph{IEEE Trans.
  Inf. Theory}, vol.~60, no.~7, pp. 4232--4265, 2014.

\end{thebibliography}


% Generated by IEEEtranTCOM.bst, version: 1.13 (2008/09/30)
\begin{thebibliography}{1}
\baselineskip 12pt
\providecommand{\url}[1]{#1}
\csname url@samestyle\endcsname
\providecommand{\newblock}{\relax}
\providecommand{\bibinfo}[2]{#2}
\providecommand{\BIBentrySTDinterwordspacing}{\spaceskip=0pt\relax}
\providecommand{\BIBentryALTinterwordstretchfactor}{4}
\providecommand{\BIBentryALTinterwordspacing}{\spaceskip=\fontdimen2\font plus
\BIBentryALTinterwordstretchfactor\fontdimen3\font minus
  \fontdimen4\font\relax}
\providecommand{\BIBforeignlanguage}[2]{{%
\expandafter\ifx\csname l@#1\endcsname\relax
\typeout{** WARNING: IEEEtran.bst: No hyphenation pattern has been}%
\typeout{** loaded for the language `#1'. Using the pattern for}%
\typeout{** the default language instead.}%
\else
\language=\csname l@#1\endcsname
\fi
#2}}
\providecommand{\BIBdecl}{\relax}
\BIBdecl

\bibitem{Popovski2014}
P.~Popovski, ``Ultra-reliable communication in 5g wireless systems,'' in
  \emph{International Conference on 5G for Ubiquitous Connectivity}, Nov. 2014,
  pp. 146--151.

\bibitem{Durisi2015}
G.~Durisi, T.~Koch, and P.~Popovski, ``Towards massive, ultra-reliable, and
  low-latency wireless: The art of sending short packets,'' pp. 1--12, 2015.

\bibitem{Polyanskiy2010b}
Y.~Polyanskiy, H.~V. Poor, and S.~Verd\'{u}, ``Channel coding rate in the
  finite blocklength regime,'' \emph{IEEE Trans. Inf. Theory}, vol.~56, no.~5,
  pp. 2307--2359, 2010.

\bibitem{spectre}
\BIBentryALTinterwordspacing
G.~Durisi, J.~Östman, Y.~Polyanskiy, I.~Tal, and W.~Yang. (2014, Dec.)
  {SPECTRE: short-packet communication toolbox, v0.2}. [Online]. Available:
  \url{https://github.com/yp-mit/spectre}
\BIBentrySTDinterwordspacing

\bibitem{Vincent2014}
\BIBentryALTinterwordspacing
V.~Y.~F. Tan and P.~Moulin, ``{Fixed error asymptotics for erasure and list
  decoding},'' \emph{arXiv}, pp. 1--18, Feb. 2013. [Online]. Available:
  \url{http://arxiv.org/abs/1301.7464v2}
\BIBentrySTDinterwordspacing

\end{thebibliography}

%\begin{biography}[{\includegraphics[width=1in,height=1.25in,clip,keepaspectratio]{mshell}}]{Michael Shell}


%\vfill



\end{document}